\documentclass[prd,preprintnumbers,twocolumn,amsmath,nofootinbib,amssymb]{revtex4}
\usepackage{graphicx,color,dcolumn,booktabs,bm}
\usepackage{longtable,lscape}
\usepackage{txfonts}
\usepackage{overpic}
\usepackage{amssymb}
\usepackage{epstopdf}
\usepackage{indentfirst}
\usepackage{feynmf}   
\usepackage{slashed}  
\usepackage{cases}
\usepackage{color}
\usepackage{float}
\usepackage{multirow}
\usepackage{ulem}
\usepackage{graphicx,color,dcolumn,booktabs,bm}
\usepackage{epsfig,dsfont,amssymb,amsmath,amsfonts,amsbsy,mathrsfs}

\graphicspath{{Figures/}} %

\usepackage{hyperref}
\hypersetup{colorlinks,citecolor=blue,anchorcolor=red,menucolor=red, linkcolor=red,filecolor=red,runcolor=red,urlcolor=blue,frenchlinks=red}

\makeatletter
\@addtoreset{equation}{section}
\makeatother

\allowdisplaybreaks

\begin{document}

\title{Surveying the mass spectra and the electromagnetic properties of the $\Xi_c^{(\prime,*)} D^{(*)}$ molecular pentaquarks}

\author{Fu-Lai Wang$^{1,2,3,5}$}
\email{wangfl2016@lzu.edu.cn}
\author{Xiang Liu$^{1,2,3,4,5}$}
\email{xiangliu@lzu.edu.cn}
\affiliation{$^1$School of Physical Science and Technology, Lanzhou University, Lanzhou 730000, China\\
$^2$Lanzhou Center for Theoretical Physics, Key Laboratory of Theoretical Physics of Gansu Province, Lanzhou University, Lanzhou 730000, China\\
$^3$Key Laboratory of Quantum Theory and Applications of MoE, Lanzhou University,
Lanzhou 730000, China\\
$^4$MoE Frontiers Science Center for Rare Isotopes, Lanzhou University, Lanzhou 730000, China\\
$^5$Research Center for Hadron and CSR Physics, Lanzhou University and Institute of Modern Physics of CAS, Lanzhou 730000, China}

\begin{abstract}
Motivated by the observed $P_{\psi s}^{\Lambda}(4459)/P_{\psi s}^{\Lambda}(4338)$ and $T_{cc}(3875)^+$ states as the $\Xi_c \bar D^{*}/\Xi_c \bar D$ and $DD^{*}$ molecular candidates, respectively, in this work we investigate the $\Xi_c^{(\prime,*)} D^{(*)}$ molecular systems. We obtain the mass spectra and the corresponding spatial wave functions of the $\Xi_c^{(\prime,*)} D^{(*)}$-type double-charm molecular pentaquark candidates with single strangeness, where we utilise the one-boson-exchange model and take into account both the $S$-$D$ wave mixing effect and the coupled channel effect. Our results show that the most promising candidates of the double-charm molecular pentaquarks with single strangeness include the $\Xi_c D$ state with $I(J^P)=0(1/2^-)$, the $\Xi_c^{\prime} D$ state with $I(J^P)=0(1/2^-)$, the $\Xi_c D^{*}$ states with $I(J^P)=0(1/2^-,3/2^-)$, the $\Xi_c^{*} D$ state with $I(J^P)=0(3/2^-)$, the $\Xi_c^{\prime} D^{*}$ states with $I(J^P)=0(1/2^-,3/2^-)$, and the $\Xi_c^{*} D^{*}$ states with $I(J^P)=0(1/2^-,3/2^-,5/2^-)$. To gain further insight into the inner structures and the properties of the isoscalar $\Xi_c^{(\prime,*)} D^{(*)}$ molecular candidates, we utilize the constituent quark model to analyze their M1 radiative decay behaviors and magnetic moments based on the obtained mass spectra and spatial wave functions, which can offer the significant information to determine their spin-parity quantum numbers and configurations in the forthcoming experiments. We suggest that our experimental colleagues search for the predicted $\Xi_c^{(\prime,*)} D^{(*)}$ molecular states.
\end{abstract}

\maketitle

\section{Introduction}\label{sec1}

The search for exotic hadronic matter is a challenging and rewarding research topic of hadron spectroscopy. In 2003, the first observation of the charmonium-like $XYZ$ state, the $X(3872)$, was reported by the Belle Collaboration \cite{Choi:2003ue}. In the last two decades, a number of new hadronic states have been observed in different experiments as summarized in reviews \cite{Liu:2013waa,Hosaka:2016pey,Chen:2016qju,Richard:2016eis,Lebed:2016hpi,Brambilla:2019esw,Liu:2019zoy,Chen:2022asf,Olsen:2017bmm,Guo:2017jvc,Meng:2022ozq}. We may notice an interesting phenomenon, where some observed new hadronic states are close to the threshold of two hadrons. It is natural to propose that these new hadronic states can be as the good candidates for the hadronic molecular states \cite{Liu:2013waa,Hosaka:2016pey,Chen:2016qju,Richard:2016eis,Lebed:2016hpi,Brambilla:2019esw,Liu:2019zoy,Chen:2022asf,Olsen:2017bmm,Guo:2017jvc,Meng:2022ozq}, which are beyond the conventional hadrons \cite{GellMann:1964nj,Zweig:1981pd}. A typical example is the discovery of three hidden-charm pentaquark states by the LHCb Collaboration, named as the $P_{\psi}^{N}(4312)$, $P_{\psi}^{N}(4440)$, and $P_{\psi}^{N}(4457)$, which exist in the $J/\psi p$ invariant mass spectrum of the $\Lambda_b^0 \to J/\psi p K^-$ decay \cite{Aaij:2019vzc}. It provides a direct evidence for the existence of the molecular-type hidden-charm  pentaquark states, which are relevant to the $\Sigma_c \bar D^{(*)}$ interaction \cite{Li:2014gra,Karliner:2015ina,Wu:2010jy,Wang:2011rga,Yang:2011wz,Wu:2012md,Chen:2015loa}.

\begin{figure}[htbp]
  \includegraphics[width=0.45\textwidth]{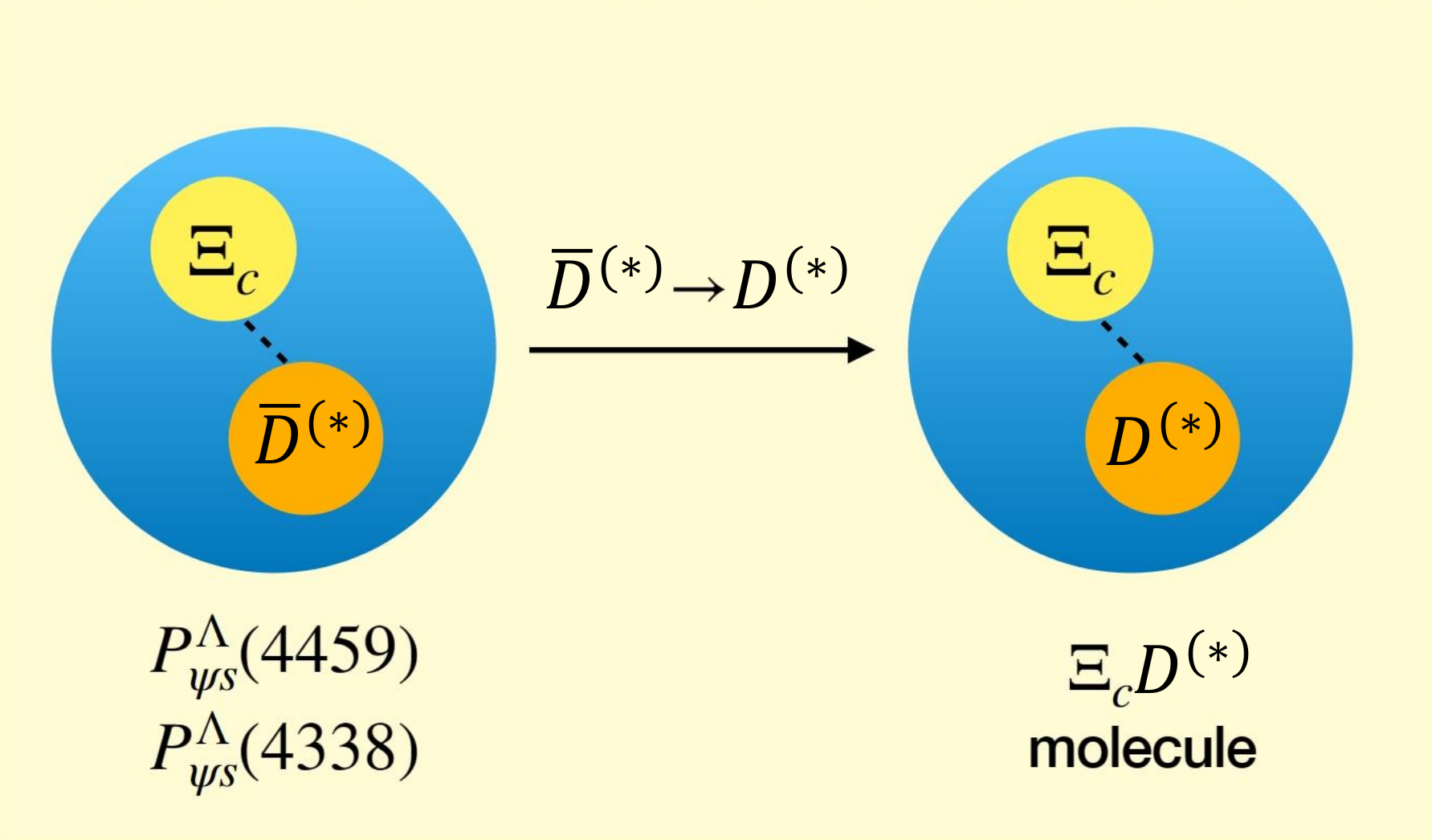}
  \caption{The $\Xi_c D^{(*)}$ molecular states are the partners of the $\Xi_c \bar D^{(*)}$ molecular states \cite{Chen:2016ryt,Wu:2010vk,Hofmann:2005sw,Anisovich:2015zqa,Wang:2015wsa,Feijoo:2015kts,Lu:2016roh,Xiao:2019gjd,Shen:2020gpw,Chen:2015sxa,Zhang:2020cdi,Wang:2019nvm,Weng:2019ynv,Paryev:2023icm,Azizi:2023foj,Wang:2022neq,Wang:2022mxy,Karliner:2022erb,Yan:2022wuz,Meng:2022wgl,Ozdem:2023htj,
Feijoo:2022rxf,Garcilazo:2022edi,Yang:2022ezl,Zhu:2022wpi,Chen:2022wkh,Ortega:2022uyu,Giachino:2022pws,Nakamura:2022jpd,Wang:2022tib,Ozdem:2022kei,Xiao:2022csb,Wang:2022gfb,Clymton:2022qlr,Chen:2022onm,Chen:2021spf,Gao:2021hmv,Ferretti:2021zis,Giron:2021fnl,Cheng:2021gca,Du:2021bgb,Chen:2021cfl,
Hu:2021nvs,Yang:2021pio,Li:2021ryu,Lu:2021irg,Zou:2021sha,Wang:2021itn,Wu:2021caw,Clymton:2021thh,Xiao:2021rgp,Ozdem:2021ugy,Zhu:2021lhd,Chen:2021tip,Azizi:2021utt,Dong:2021juy,Liu:2020hcv,Wang:2020eep,Peng:2020hql,Chen:2020uif,Peng:2019wys,Chen:2020opr,Chen:2020kco,Burns:2022uha} by substituting the $\bar D^{(*)}$ meson with the $D^{(*)}$ meson.}\label{Extension1}
\end{figure}

Subsequently, two hidden-charm pentaquark states with strangeness, the $P_{\psi s}^{\Lambda}(4459)$ \cite{LHCb:2020jpq} and $P_{\psi s}^{\Lambda}(4338)$ \cite{LHCb:2022ogu}, were announced by LHCb. This is a good opportunity to briefly present the research status of them. In 2020, the $P_{\psi s}^{\Lambda}(4459)$ \cite{LHCb:2020jpq} was found by analyzing the $\Xi_b^-\to J/\psi \Lambda K^-$ process, which can be regarded as the $\Xi_c \bar{D}^*$ molecular candidate \cite{Chen:2016ryt,Wu:2010vk,Hofmann:2005sw,Anisovich:2015zqa,Wang:2015wsa,Feijoo:2015kts,Lu:2016roh,Xiao:2019gjd,Shen:2020gpw,Chen:2015sxa,Zhang:2020cdi,Wang:2019nvm,Weng:2019ynv,Paryev:2023icm,Azizi:2023foj,Wang:2022neq,Wang:2022mxy,Karliner:2022erb,Yan:2022wuz,Meng:2022wgl,Ozdem:2023htj,
Feijoo:2022rxf,Garcilazo:2022edi,Yang:2022ezl,Zhu:2022wpi,Chen:2022wkh,Ortega:2022uyu,Giachino:2022pws,Nakamura:2022jpd,Wang:2022tib,Ozdem:2022kei,Xiao:2022csb,Wang:2022gfb,Clymton:2022qlr,Chen:2022onm,Chen:2021spf,Gao:2021hmv,Ferretti:2021zis,Giron:2021fnl,Cheng:2021gca,Du:2021bgb,Chen:2021cfl,
Hu:2021nvs,Yang:2021pio,Li:2021ryu,Lu:2021irg,Zou:2021sha,Wang:2021itn,Wu:2021caw,Clymton:2021thh,Xiao:2021rgp,Ozdem:2021ugy,Zhu:2021lhd,Chen:2021tip,Azizi:2021utt,Dong:2021juy,Liu:2020hcv,Wang:2020eep,Peng:2020hql,Chen:2020uif,Peng:2019wys,Chen:2020opr,Chen:2020kco,Burns:2022uha}. Later, LHCb found the $P_{\psi s}^{\Lambda}(4338)$ \cite{LHCb:2022ogu} in the $J/\psi\Lambda$ invariant mass spectrum of the $B^-\to J/\psi\Lambda\bar{p}$ weak decay, which can be interpreted as the $\Xi_c \bar{D}$ molecular candidate \cite{Chen:2016ryt,Wu:2010vk,Hofmann:2005sw,Anisovich:2015zqa,Wang:2015wsa,Feijoo:2015kts,Lu:2016roh,Xiao:2019gjd,Shen:2020gpw,Chen:2015sxa,Zhang:2020cdi,Wang:2019nvm,Weng:2019ynv,Paryev:2023icm,Azizi:2023foj,Wang:2022neq,Wang:2022mxy,Karliner:2022erb,Yan:2022wuz,Meng:2022wgl,Ozdem:2023htj,
Feijoo:2022rxf,Garcilazo:2022edi,Yang:2022ezl,Zhu:2022wpi,Chen:2022wkh,Ortega:2022uyu,Giachino:2022pws,Nakamura:2022jpd,Wang:2022tib,Ozdem:2022kei,Xiao:2022csb,Wang:2022gfb,Clymton:2022qlr,Chen:2022onm,Chen:2021spf,Gao:2021hmv,Ferretti:2021zis,Giron:2021fnl,Cheng:2021gca,Du:2021bgb,Chen:2021cfl,
Hu:2021nvs,Yang:2021pio,Li:2021ryu,Lu:2021irg,Zou:2021sha,Wang:2021itn,Wu:2021caw,Clymton:2021thh,Xiao:2021rgp,Ozdem:2021ugy,Zhu:2021lhd,Chen:2021tip,Azizi:2021utt,Dong:2021juy,Liu:2020hcv,Wang:2020eep,Peng:2020hql,Chen:2020uif,Peng:2019wys,Chen:2020opr,Chen:2020kco,Burns:2022uha}. Interestingly, the mass gap between the $P_{\psi s}^{\Lambda}(4459)$ and $P_{\psi s}^{\Lambda}(4338)$ states is similar to that of these observed $p_{\psi}^N$ states \cite{Karliner:2022erb,Wang:2022mxy}, showing the possibility of the existence of the double peak structures located just below the threshold of the $\Xi_c\bar D^*$ channel for the $P_{\psi s}^\Lambda(4459)$ \cite{Karliner:2022erb,Wang:2022mxy}. It awaits future testing by more accurate data. Having these $P_{\psi s}^{\Lambda}(4459)$ and $P_{\psi s}^{\Lambda}(4338)$ states, we can make an extension as shown in Fig. \ref{Extension1}, where $\bar{D}^{(*)}$ is replaced by ${D}^{(*)}$. This treatment makes us interested in exploring
the $\Xi_c D^{(*)}$ molecular states, which belong to the typical double-charm molecular pentaquark systems.

\begin{figure}[htbp]
  \includegraphics[width=0.45\textwidth]{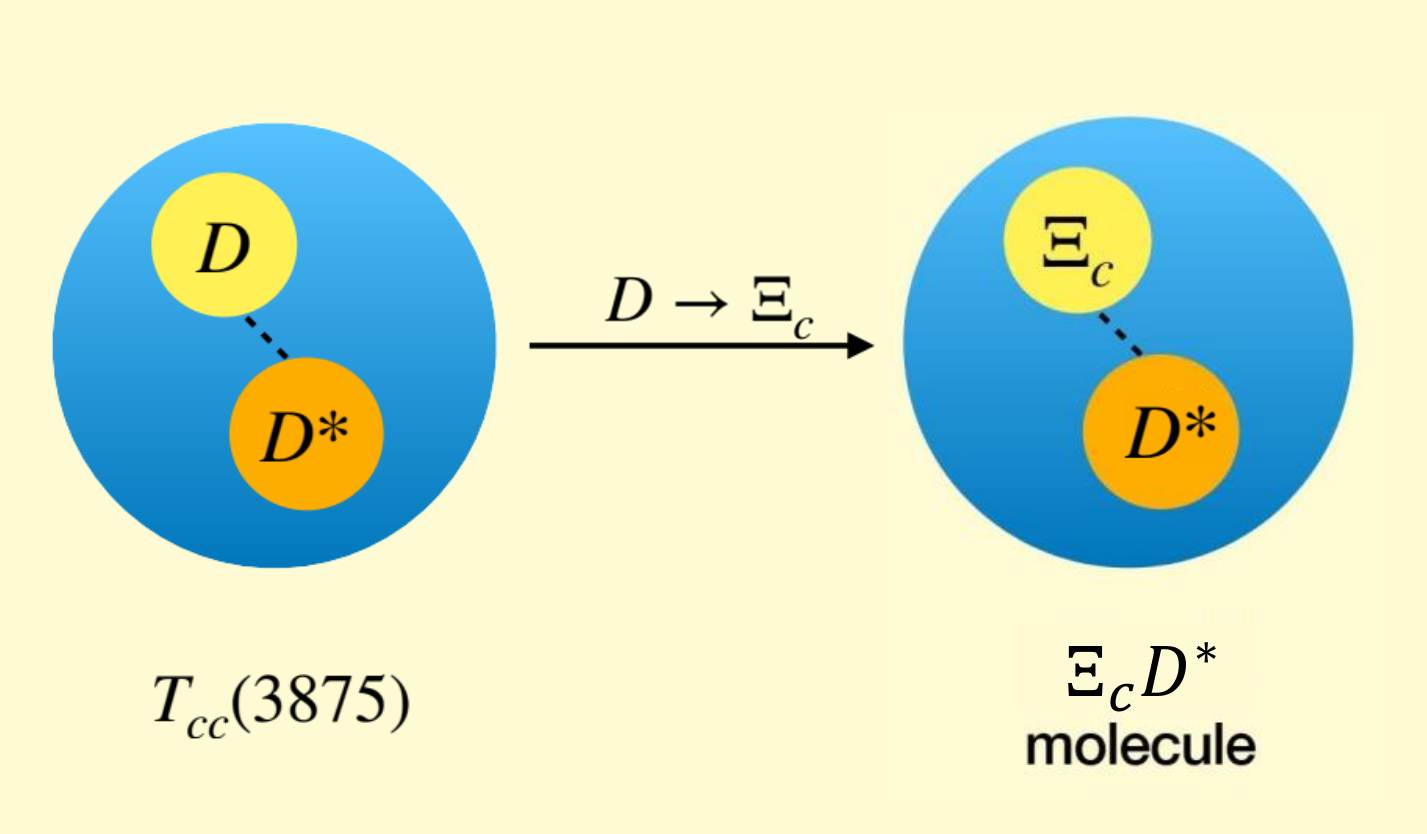}
  \caption{The investigation of the $\Xi_c D^{*}$ molecular states is a logical extension of the observed  $T_{cc}(3875)^+$ state \cite{LHCb:2021vvq} as the $DD^{*}$  molecular tetraquark candidate \cite{Manohar:1992nd,Ericson:1993wy,Tornqvist:1993ng,Janc:2004qn,Ding:2009vj,Molina:2010tx,Ding:2020dio,Li:2012ss,Xu:2017tsr,Liu:2019stu,Ohkoda:2012hv,Tang:2019nwv,Xin:2021wcr}.}\label{Extension2}
\end{figure}

In fact, the discussed double-charm molecular $\Xi_c D^{*}$ pentaquarks are closely related to the $T_{cc}(3875)^+$ observed in the $D^0D^0\pi^+$ invariant mass spectrum produced by the $pp$ collisions \cite{LHCb:2021vvq}, which can be explained as the $D D^{*}$ molecular tetraquark candidate \cite{Manohar:1992nd,Ericson:1993wy,Tornqvist:1993ng,Janc:2004qn,Ding:2009vj,Molina:2010tx,Ding:2020dio,Li:2012ss,Xu:2017tsr,Liu:2019stu,Ohkoda:2012hv,Tang:2019nwv,Xin:2021wcr}\footnote{The flavor representation of the light antiquark in the $D$ is $\bar 3_F$, while the flavor representations of the diquark in the $\Xi_c$ and $\Xi_c^{\prime(*)}$ are $\bar 3_F$ and $6_F$, respectively. Thus, $\Xi_c$ should be much more similar to $D$, but $\Xi_c^{\prime(*)}$ is not.}. As illustrated in
Fig. \ref{Extension2}, the discussed $\Xi_c D^{*}$ molecular pentaquarks are a logical extension of the $T_{cc}(3875)^+$ state due to the replacement of $D\to \Xi_c$. Obviously, the study of the $\Xi_c D^{*}$ double-charm molecular pentaquarks enhances our understanding of the double-heavy exotic multiquarks \cite{Chen:2021kad,Yalikun:2023waw,Wang:2023aob}. The present work may encourage our experimental colleagues to have ambitions to find out the predicted  $\Xi_c D^{*}$ molecular pentaquarks.

In this paper, we study the mass spectra and the corresponding spatial wave functions of the $\Xi_c^{(\prime,*)} D^{(*)}$-type double-charm molecular pentaquark candidates with single strangeness, which are important not only for experimentally searching for these molecular pentaquarks but also for theoretically analyzing their properties. In our concrete analysis, the interactions of the $\Xi_c^{(\prime,*)} D^{(*)}$ systems can be investigated through the one-boson-exchange (OBE) model \cite{Chen:2016qju, Liu:2019zoy} and take into account both the $S$-$D$ wave mixing effect and the coupled channel effect, and the coupled channel Schr$\ddot{\rm o}$dinger equation is then employed to analyze their bound state properties including the binding energies and the spatial wave functions, which is the successful approach to reproduce the bound state properties of the deuteron as the molecular state consisting of the neutron and the proton \cite{Machleidt:1987hj,Epelbaum:2008ga,Esposito:2014rxa,Chen:2016qju,Tornqvist:1993ng,Tornqvist:1993vu,Wang:2019nwt,Chen:2017jjn}. After that, we investigate the M1 radiative decay behaviors and the magnetic moments of the isoscalar $\Xi_c^{(\prime,*)} D^{(*)}$ molecular pentaquarks based on the obtained mass spectra and spatial wave functions, which can provide the deeper insights into their inner structures and properties. In our calculations, the constituent quark model is employed, which is the widely accepted method for exploring the electromagnetic properties of the hadronic states \cite{Liu:2003ab,Huang:2004tn,Zhu:2004xa,Haghpayma:2006hu,Wang:2016dzu,Deng:2021gnb,Gao:2021hmv,Zhou:2022gra,Wang:2022tib,Li:2021ryu,Schlumpf:1992vq,Schlumpf:1993rm,Cheng:1997kr,Ha:1998gf,Ramalho:2009gk,Girdhar:2015gsa,Menapara:2022ksj,Mutuk:2021epz,Menapara:2021vug,Menapara:2021dzi,Gandhi:2018lez,Dahiya:2018ahb,Kaur:2016kan,Thakkar:2016sog,Shah:2016vmd,Dhir:2013nka,Sharma:2012jqz,Majethiya:2011ry,Sharma:2010vv,Dhir:2009ax,Simonis:2018rld,Ghalenovi:2014swa,Kumar:2005ei,Rahmani:2020pol,Hazra:2021lpa,Gandhi:2019bju,Majethiya:2009vx,Shah:2016nxi,Shah:2018bnr,Ghalenovi:2018fxh,Wang:2022nqs,Mohan:2022sxm,An:2022qpt,Kakadiya:2022pin,Wu:2022gie,Wang:2023bek,Wang:2023aob}. In addition, other approaches such as the light-cone QCD sum rules are often used to discuss the magnetic moments of the molecular pentaquark states \cite{Xu:2020flp,Ozdem:2022kei,Ozdem:2023htj,Ozdem:2021ugy}.

The structure of this paper is as follows: In Sec. \ref{sec2}, we utilise the OBE model to derive the interactions for the $\Xi_c^{(\prime,*)} D^{(*)}$ systems. Following this, we obtain the mass spectra and the corresponding spatial wave functions of the $\Xi_c^{(\prime,*)} D^{(*)}$-type double-charm molecular pentaquark candidates with single strangeness. In Sec. \ref{sec3}, we investigate the M1 radiative decay behaviors and the magnetic moments of the isoscalar $\Xi_c^{(\prime,*)} D^{(*)}$ molecular pentaquarks based on the obtained mass spectra and spatial wave functions using the constituent quark model. In Sec. \ref{sec4}, a summary is presented.

\section{The mass spectra and the corresponding spatial wave functions}\label{sec2}

In this section, we derive the OBE effective potentials of the $\Xi_c^{(\prime,*)} D^{(*)}$ systems, and then their bound state properties comprising of the binding energies $E$ and the corresponding spatial wave functions $\phi(r)$ can be discussed by solving the coupled channel Schr\"{o}dinger equation, which is important for establishing the mass spectra and investigating the properties of the $\Xi_c^{(\prime,*)} D^{(*)}$ molecular pentaquarks.

\subsection{The OBE effective potentials of the $\Xi_c^{(\prime,*)} D^{(*)}$ systems}

In this study, we utilise the OBE model to explore the interactions of the $\Xi_c^{(\prime,*)} D^{(*)}$ systems, which include bosons such as the light scalar meson $\sigma$, the light pseudoscalar mesons $\pi$ and $\eta$, as well as the light vector mesons $\rho$ and $\omega$ \cite{Chen:2016qju, Liu:2019zoy}. In the concrete calculations, the effective Lagrangian approach is utilized. Thus, it is crucial to construct the effective Lagrangians describing the interactions among the heavy hadrons $\Xi_c^{(\prime,*)}/{D}^{(*)}$ and the light scalar, pseudoscalar, and vector mesons $\mathcal{E}$.

\subsubsection{The effective Lagrangians concerning the interactions between the heavy hadrons $\Xi_c^{(\prime,*)}/{D}^{(*)}$ and the light mesons $\mathcal{E}$}

For the $S$-wave single-charmed baryons in the $6_F$ flavour representation $\mathcal{B}_6$ and $\mathcal{B}_{6}^*$, it is necessary to construct the superfield $\mathcal{S}_{\mu}$ according to the heavy quark spin symmetry \cite{Wise:1992hn,Casalbuoni:1992gi,Casalbuoni:1996pg,Yan:1992gz,Bando:1987br,Harada:2003jx,Chen:2017xat}, i.e.,
\begin{eqnarray}
\mathcal{S}_{\mu}&=&-\sqrt{\frac{1}{3}}(\gamma_{\mu}+v_{\mu})\gamma^5\mathcal{B}_6+\mathcal{B}_{6\mu}^*.
\end{eqnarray}
Furthermore, the superfield $H^{({Q})}_a$ can be formed by utilizing the $S$-wave charmed mesons ${D}$ with $J^P=0^-$ and ${D}^{*}$ with $J^P=1^-$, which can be expressed in the following manner \cite{Ding:2008gr}
\begin{eqnarray}
H^{(Q)}_a&=&\frac{1+{v}\!\!\!\slash}{2}\left(D^{*\mu}_a\gamma_{\mu}-D_a\gamma_5\right).
\end{eqnarray}
Here, the four-velocity is represented as $v_{\mu}$ and has the value $v_{\mu}=(1,\bm{0})$ when making the non-relativistic approximation. In addition, the associated conjugate fields are $\bar{\mathcal{S}}_{\mu}=\mathcal{S}_{\mu}^\dag\gamma^0$ and $\bar{H}^{(Q)}_a=\gamma^0H_a^{(Q)\dag}\gamma^0$
\cite{Wise:1992hn,Casalbuoni:1992gi,Casalbuoni:1996pg,Yan:1992gz,Bando:1987br,Harada:2003jx,Chen:2017xat,Ding:2008gr}. For the previously defined superfields $\mathcal{S}_{\mu}$ and $H^{({Q})}_a$, the normalization relations for the single-charmed baryons $\mathcal{B}$ and $\mathcal{B}^{*}$ are $\langle 0|\mathcal{B}|cqq\left({1}/{2}^+\right)\rangle = \sqrt{2m_{\mathcal{B}}}{\left(\chi_{\frac{1}{2}m},\frac{\bm{\sigma}\cdot\bm{p}}{2m_{\mathcal{B}}}\chi_{\frac{1}{2}m}\right)^T}$ and
$\langle 0|\mathcal{B}^{*\mu}|cqq\left({3}/{2}^+\right)\rangle =\sqrt{2m_{\mathcal{B}^*}}\left(\Phi_{\frac{3}{2}m}^{\mu},\frac{\bm{\sigma}\cdot\bm{p}}{2m_{\mathcal{B}^*}}\Phi_{\frac{3}{2}m}^{\mu}\right)^T$, and the normalization relations for the charmed mesons $D$ and $D^{*}$ are $\langle 0|D|c\bar{q}\left(0^-\right)\rangle=\sqrt{m_{D}}$
and
$\langle 0|D^{*\mu}|c\bar{q}\left(1^-\right)\rangle=\sqrt{m_{D^*}}\epsilon^\mu$ \cite{Wise:1992hn,Casalbuoni:1992gi,Casalbuoni:1996pg,Yan:1992gz,Bando:1987br,Harada:2003jx,Chen:2017xat,Ding:2008gr}. Here, the mass of the hadronic state $i$ is $m_{i}$, the Pauli matrix of the charmed baryon $\mathcal{B}^{(*)}$ is $\bm{\sigma}$, and the momentum of the charmed baryon $\mathcal{B}^{(*)}$ is $\bm{p}$. The polarization vector $\epsilon_{m}^{\mu}\,(m=0,\,\pm1)$ assumes the values of $\epsilon_{\pm}^{\mu}= \left(0,\,\pm1,\,i,\,0\right)/\sqrt{2}$ and $\epsilon_{0}^{\mu}= \left(0,0,0,-1\right)$ within the static limit. The spin wave function describing the baryon with the spin of $S={1}/{2}$ is represented by $\chi_{\frac{1}{2}m}$, the polarization tensor $\Phi_{\frac{3}{2}m}^{\mu}$ represents the spin wave function of the baryon with the spin of $S={3}/{2}$ and can be represented as $\Phi_{\frac{3}{2}m}^{\mu}=\sum_{m_1,m_2}C^{\frac{3}{2}m}_{\frac{1}{2}m_1,1m_2}\chi_{\frac{1}{2}m_1}\epsilon_{m_2}^{\mu}$, where the constant $C^{ef}_{ab,cd}$ is the Clebsch-Gordan coefficient.

Utilizing the restrictions of the heavy quark symmetry, the chiral symmetry, and the hidden local symmetry \cite{Casalbuoni:1992gi,Casalbuoni:1996pg,Yan:1992gz,Harada:2003jx,Bando:1987br}, the effective Lagrangians for the interactions among the heavy hadronic states $\Xi_c^{(\prime,*)}/{D}^{(*)}$ and the light scalar, pseudoscalar, and vector mesons $\mathcal{E}$ were constructed in Refs. \cite{Wise:1992hn,Casalbuoni:1992gi,Casalbuoni:1996pg,Yan:1992gz,Bando:1987br,Harada:2003jx,Chen:2017xat,Ding:2008gr}, which are explicitly written as
\begin{eqnarray}
\mathcal{L}_{\mathcal{B}_{\bar{3}}\mathcal{B}_{\bar{3}}\mathcal{E}} &=& l_B\langle\bar{\mathcal{B}}_{\bar{3}}\sigma\mathcal{B}_{\bar{3}}\rangle
          +i\beta_B\langle\bar{\mathcal{B}}_{\bar{3}}v^{\mu}(\mathcal{V}_{\mu}
          -\rho_{\mu})\mathcal{B}_{\bar{3}}\rangle,\\
\mathcal{L}_{\mathcal{S}\mathcal{S}\mathcal{E}} &=& l_S\langle\bar{\mathcal{S}}_{\mu}\sigma\mathcal{S}^{\mu}\rangle -\frac{3}{2}g_1\varepsilon^{\mu\nu\lambda\kappa}v_{\kappa}\langle\bar{\mathcal{S}}_{\mu}\mathcal{A}_{\nu}
\mathcal{S}_{\lambda}\rangle\nonumber\\
    &&+i\beta_{S}\langle\bar{\mathcal{S}}_{\mu}v_{\alpha}\left(\mathcal{V}^{\alpha}
    -\rho^{\alpha}\right) \mathcal{S}^{\mu}\rangle +\lambda_S\langle\bar{\mathcal{S}}_{\mu}F^{\mu\nu}(\rho)\mathcal{S}_{\nu}\rangle,\nonumber\\\\
\mathcal{L}_{\mathcal{B}_{\bar{3}}\mathcal{S}\mathcal{E}}&=&ig_4\langle\bar{\mathcal{S}^{\mu}}
\mathcal{A}_{\mu}\mathcal{B}_{\bar{3}}\rangle+i\lambda_I\varepsilon^{\mu\nu\lambda
\kappa}v_{\mu}\langle \bar{\mathcal{S}}_{\nu}F_{\lambda\kappa}\mathcal{B}_{\bar{3}}\rangle+h.c.,\nonumber\\\\
\mathcal{L}_{HH\mathcal{E}}&=&g_{\sigma}\left\langle H^{(Q)}_a\sigma\bar{H}^{(Q)}_a\right\rangle+ig\left\langle H^{(Q)}_b{\mathcal A}\!\!\!\slash_{ba}\gamma_5\bar{H}^{\,({Q})}_a\right\rangle\nonumber\\
  &&+i\beta\left\langle H^{(Q)}_b v^{\mu}({\mathcal V}_{\mu}-\rho_{\mu})_{ba}\bar{H}^{\,(Q)}_a\right\rangle\nonumber\\
  &&+i\lambda\left\langle H^{(Q)}_b \sigma^{\mu\nu}F_{\mu\nu}(\rho)_{ba}\bar{H}^{\,(Q)}_a\right\rangle.
\end{eqnarray}
In the effective Lagrangians presented above, we identify the $S$-wave single-charmed baryons in the $\bar{3}_F$ flavor representation and the $6_F$ flavor representation as $\mathcal{B}_{\bar{3}}$ and $\mathcal{B}_6^{(*)}$, which are  \cite{Wise:1992hn,Casalbuoni:1992gi,Casalbuoni:1996pg,Yan:1992gz,Bando:1987br,Harada:2003jx,Chen:2017xat}
\begin{eqnarray}
\mathcal{B}_{\bar{3}} = \left(\begin{array}{ccc}
        0    &\Lambda_c^+      &\Xi_c^+\\
        -\Lambda_c^+       &0      &\Xi_c^0\\
        -\Xi_c^+      &-\Xi_c^0     &0
\end{array}\right),
\mathcal{B}_6^{(*)} = \left(\begin{array}{ccc}
         \Sigma_c^{{(*)}++}                  &\frac{\Sigma_c^{{(*)}+}}{\sqrt{2}}     &\frac{\Xi_c^{\prime(*)+}}{\sqrt{2}}\\
         \frac{\Sigma_c^{{(*)}+}}{\sqrt{2}}      &\Sigma_c^{{(*)}0}    &\frac{\Xi_c^{\prime(*)0}}{\sqrt{2}}\\
         \frac{\Xi_c^{\prime(*)+}}{\sqrt{2}}    &\frac{\Xi_c^{\prime(*)0}}{\sqrt{2}}      &\Omega_c^{(*)0}
\end{array}\right),
\end{eqnarray}
respectively. $\mathcal{A}_\mu$ and ${\mathcal V}_{\mu}$ are the axial current and the vector current, which can be written as
\begin{eqnarray}
{\mathcal A}_{\mu}=\frac{1}{2}\left(\xi^{\dagger}\partial_{\mu}\xi-\xi\partial_{\mu}\xi^{\dagger}\right),~~
{\mathcal V}_{\mu}=\frac{1}{2}\left(\xi^{\dagger}\partial_{\mu}\xi+\xi\partial_{\mu}\xi^{\dagger}\right),
\end{eqnarray}
respectively. $\xi=e^{i\mathbb{P}/f_\pi}$ is the pseudo-Goldstone meson field with $f_{\pi}$ being equal to $132~{\rm MeV}$, and the light pseudoscalar meson matrix $\mathbb{P}$ is \cite{Wise:1992hn,Casalbuoni:1992gi,Casalbuoni:1996pg,Yan:1992gz,Bando:1987br,Harada:2003jx,Chen:2017xat,Ding:2008gr}
\begin{eqnarray}
{\mathbb{P}} &=& {\left(\begin{array}{ccc}
       \frac{\pi^0}{\sqrt{2}}+\frac{\eta}{\sqrt{6}} &\pi^+ &K^+\\
       \pi^-       &-\frac{\pi^0}{\sqrt{2}}+\frac{\eta}{\sqrt{6}} &K^0\\
       K^-         &\bar K^0   &-\sqrt{\frac{2}{3}} \eta     \end{array}\right)}.
\end{eqnarray}
Furthermore, the vector meson field and the vector meson field strength tensor are defined as $\rho_{\mu}$ and $F_{\mu\nu}$, which are explicitly written as
$\rho_{\mu}=i{g_V}\mathbb{V}_{\mu}/{\sqrt{2}}$ and
$F_{\mu\nu}=\partial_{\mu}\rho_{\nu}-\partial_{\nu}\rho_{\mu}+[\rho_{\mu},\rho_{\nu}]$,
respectively. Here, the light vector meson matrix $\mathbb{V}_{\mu}$ is \cite{Wise:1992hn,Casalbuoni:1992gi,Casalbuoni:1996pg,Yan:1992gz,Bando:1987br,Harada:2003jx,Chen:2017xat,Ding:2008gr}
\begin{eqnarray}
{\mathbb{V}}_{\mu} &=& {\left(\begin{array}{ccc}
       \frac{\rho^0}{\sqrt{2}}+\frac{\omega}{\sqrt{2}} &\rho^+ &K^{*+}\\
       \rho^-       &-\frac{\rho^0}{\sqrt{2}}+\frac{\omega}{\sqrt{2}} &K^{*0}\\
       K^{*-}         &\bar K^{*0}   & \phi     \end{array}\right)}_{\mu}.
\end{eqnarray}

Following the preceding discussions, we can derive the specific effective Lagrangians illustrating the interactions between the heavy hadrons $\Xi_c^{(\prime,*)}/{D}^{(*)}$ and the light mesons $\mathcal{E}$, which can be achieved by expanding the effective Lagrangians constructed above to the leading order of the pseudo-Goldstone meson field $\xi$ as explained below
\begin{eqnarray}
\mathcal{L}_{\mathcal{B}_{\bar{3}}\mathcal{B}_{\bar{3}}\sigma}&=& l_B\langle \bar{\mathcal{B}}_{\bar{3}}\sigma\mathcal{B}_{\bar{3}}\rangle,\\
\mathcal{L}_{\mathcal{B}_{6}^{(*)}\mathcal{B}_{6}^{(*)}\sigma}&=&-l_S\langle\bar{\mathcal{B}}_6\sigma\mathcal{B}_6\rangle+l_S\langle
\bar{\mathcal{B}}_{6\mu}^{*}
\sigma\mathcal{B}_6^{*\mu}\rangle\nonumber\\
    &&-\frac{l_S}{\sqrt{3}}\langle\bar{\mathcal{B}}_{6\mu}^{*}\sigma
    \left(\gamma^{\mu}+v^{\mu}\right)\gamma^5\mathcal{B}_6\rangle+h.c.,\\
\mathcal{L}_{HH \sigma} &=&-2g_{\sigma}{D}_a \sigma {D}_a^{\dag}+ 2g_{\sigma} {D}_{a\mu}^* \sigma {D}_a^{*\mu\dag},\\
\mathcal{L}_{\mathcal{B}_6^{(*)}\mathcal{B}_6^{(*)}\mathbb{P}}&=&i\frac{g_1}{2f_{\pi}}\varepsilon^{\mu\nu\lambda\kappa}v_{\kappa}\langle\bar{\mathcal{B}}_6
\gamma_{\mu}\gamma_{\lambda}\partial_{\nu}\mathbb{P}\mathcal{B}_6\rangle\nonumber\\
    &&-i\frac{3g_1}{2f_{\pi}}\varepsilon^{\mu\nu\lambda\kappa}v_{\kappa}
    \langle\bar{\mathcal{B}}_{6\mu}^{*}\partial_{\nu}\mathbb{P}\mathcal{B}_{6\lambda}^*\rangle\nonumber\\
    &&+i\frac{\sqrt{3}g_1}{2f_{\pi}}v_{\kappa}\varepsilon^{\mu\nu\lambda\kappa}
    \langle\bar{\mathcal{B}}_{6\mu}^*\partial_{\nu}\mathbb{P}{\gamma_{\lambda}\gamma^5}
      \mathcal{B}_6\rangle+h.c.,\\
\mathcal{L}_{\mathcal{B}_{\bar{3}}\mathcal{B}_6^{(*)}\mathbb{P}} &=& -\sqrt{\frac{1}{3}}\frac{g_4}{f_{\pi}}\langle\bar{\mathcal{B}}_6\gamma^5\left(\gamma^{\mu}
+v^{\mu}\right)\partial_{\mu}\mathbb{P}\mathcal{B}_{\bar{3}}\rangle\nonumber\\
    &&-\frac{g_4}{f_{\pi}}\langle\bar{\mathcal{B}}_{6\mu}^*\partial^{\mu} \mathbb{P}\mathcal{B}_{\bar{3}}\rangle+h.c.,\\
\mathcal{L}_{HH\mathbb{P}}&=&-\frac{2ig}{f_{\pi}}v^{\alpha}\varepsilon_{\alpha\mu\nu\lambda}D_b^{*\mu}D_a^{*\lambda\dag}\partial^{\nu}{\mathbb{P}}_{ba}\nonumber\\
    &&-\frac{2g}{f_{\pi}}(D_b^{*\mu}D_a^{\dag}+D_bD_a^{*\mu\dag})\partial_{\mu}{\mathbb{P}}_{ba},\\
\mathcal{L}_{\mathcal{B}_{\bar{3}}\mathcal{B}_{\bar{3}}\mathbb{V}}&=&
\frac{1}{\sqrt{2}}\beta_Bg_V\langle\bar{\mathcal{B}}_{\bar{3}}v\cdot\mathbb{V}
\mathcal{B}_{\bar{3}}\rangle,\\
\mathcal{L}_{\mathcal{B}_6^{(*)}\mathcal{B}_6^{(*)}\mathbb{V}}&=&
-\frac{\beta_Sg_V}{\sqrt{2}}\langle\bar{\mathcal{B}}_6v\cdot\mathbb{V}
\mathcal{B}_6\rangle+\frac{\beta_Sg_V}{\sqrt{2}}\langle\bar{\mathcal{B}}_{6\mu}^*v\cdot {V}\mathcal{B}_6^{*\mu}\rangle\nonumber\\
    &&-i\frac{\lambda_S g_V}{3\sqrt{2}}\langle\bar{\mathcal{B}}_6\gamma_{\mu}\gamma_{\nu}
    \left(\partial^{\mu}\mathbb{V}^{\nu}-\partial^{\nu}\mathbb{V}^{\mu}\right)
    \mathcal{B}_6\rangle\nonumber\\
    &&+i\frac{\lambda_Sg_V}{\sqrt{2}}\langle\bar{\mathcal{B}}_{6\mu}^*
    \left(\partial^{\mu}\mathbb{V}^{\nu}-\partial^{\nu}\mathbb{V}^{\mu}\right)
    \mathcal{B}_{6\nu}^*\rangle\nonumber\\
    &&-i\frac{\lambda_Sg_V}{\sqrt{6}}\langle\bar{\mathcal{B}}_{6\mu}^*
    \left(\partial^{\mu}\mathbb{V}^{\nu}-\partial^{\nu}\mathbb{V}^{\mu}\right)
    \left(\gamma_{\nu}+v_{\nu}\right)\gamma^5\mathcal{B}_6\rangle\nonumber\\
    &&-\frac{\beta_Sg_V}{\sqrt{6}}\langle\bar{\mathcal{B}}_{6\mu}^*v\cdot \mathbb{V}\left(\gamma^{\mu}+v^{\mu}\right)\gamma^5\mathcal{B}_6\rangle+h.c.,\\
\mathcal{L}_{\mathcal{B}_{\bar{3}}\mathcal{B}_6^{(*)}\mathbb{V}} &=&
-\frac{\lambda_Ig_V}{\sqrt{2}}\varepsilon^{\mu\nu\lambda\kappa}v_{\mu}\langle \bar{\mathcal{B}}_{6\nu}^*\left(\partial_{\lambda}\mathbb{V}_{\kappa}
-\partial_{\kappa}\mathbb{V}_{\lambda}\right)
          \mathcal{B}_{\bar{3}}\rangle\nonumber\\
    &&-\frac{\lambda_Ig_V}{\sqrt{6}}\varepsilon^{\mu\nu\lambda\kappa}v_{\mu}\langle \bar{\mathcal{B}}_6\gamma^5\gamma_{\nu}
        \left(\partial_{\lambda}\mathbb{V}_{\kappa}-\partial_{\kappa} \mathbb{V}_{\lambda}\right)\mathcal{B}_{\bar{3}}\rangle+h.c.,\nonumber\\\\
\mathcal{L}_{HH\mathbb{V}} &=&-\sqrt{2}\beta g_V D_b D_a^{\dag} v\cdot\mathbb{V}_{ba}+\sqrt{2}\beta g_V D_{b\mu}^* D_a^{*\mu\dag}v\cdot\mathbb{V}_{ba}\nonumber\\
    &&-2\sqrt{2}i\lambda g_V D_b^{*\mu}D_a^{*\nu\dag}\left(\partial_{\mu}\mathbb{V}_{\nu}-\partial_{\nu}\mathbb{V}_{\mu}\right)_{ba}\nonumber\\
    &&-2\sqrt{2}\lambda g_V v^{\lambda}\varepsilon_{\lambda\mu\alpha\beta}(D_bD_a^{*\mu\dag}+D_b^{*\mu}D_a^{\dag})\partial^{\alpha}\mathbb{V}^{\beta}_{ba}.
\end{eqnarray}

The above effective Lagrangians contain several coupling constants to depict the interaction strengths between the heavy hadrons $\Xi_c^{(\prime,*)}/{D}^{(*)}$ and the light mesons $\mathcal{E}$, and these values can be constrained by matching the experimental data or taking the theoretical models. In addition, the phase factors between the relevant coupling constants can then be determined by applying the quark model \cite{Riska:2000gd}. In the present work, the values of $l_B=-3.65$, $l_S=6.20$, $g_{\sigma}=0.76$, $g_1=0.94$, $g_4=1.06$, $g=-0.59$, $\beta_B g_{V}=-6.00$, $\beta_S g_{V}=10.14$, $\beta g_{V}=-5.25$, $\lambda_Ig_V=-6.80~\rm {GeV^{-1}}$, $\lambda_Sg_V=19.20~\rm {GeV^{-1}}$, and $\lambda g_{V}=-3.27 ~\rm {GeV^{-1}}$ are taken \cite{Wang:2022mxy,Chen:2017xat,Chen:2019asm,Chen:2020kco,Wang:2020bjt,Wang:2019nwt,Chen:2018pzd,Wang:2021hql,Wang:2021ajy,Wang:2021yld,Wang:2021aql,Yang:2021sue,Wang:2020dya,Wang:2023ftp,Yalikun:2023waw,Wang:2023aob}. For the masses of the mesons and the baryons, we utilise $m_\sigma=600.00$ MeV, $m_\pi=137.27$ MeV, $m_\eta=547.86$ MeV, $m_\rho=775.26$ MeV, $m_\omega=782.66$ MeV, $m_{\Xi_{c}^{+}}=2467.71$ MeV, $m_{\Xi_{c}^{0}}=2470.44$ MeV, $m_{\Xi_{c}^{\prime+}}=2578.20$ MeV,  $m_{\Xi_{c}^{\prime 0}}=2578.70$ MeV, $m_{\Xi_{c}^{*+}}=2645.10$ MeV, $m_{\Xi_{c}^{*0}}=2645.16$ MeV, $m_{D^{+}}=1869.66$ MeV, $m_{D^{0}}=1864.84$ MeV, $m_{D^{*+}}=2006.85$ MeV, and $m_{D^{*0}}=2010.26$ MeV \cite{ParticleDataGroup:2022pth}.

\subsubsection{The OBE effective potentials of the $\Xi_c^{(\prime,*)} D^{(*)}$ systems}

By referring to the obtained effective Lagrangians \cite{Wise:1992hn,Casalbuoni:1992gi,Casalbuoni:1996pg,Yan:1992gz,Bando:1987br,Harada:2003jx,Chen:2017xat,Ding:2008gr}, we can determine the scattering amplitude $\mathcal{M}^{\Xi_{c}^{(\prime, *)}{D}^{(*)}\to \Xi_{c}^{(\prime, *)}{D}^{(*)}}(\bm{q})$ for the $\Xi_{c}^{(\prime,*)}{D}^{(*)}\to \Xi_{c}^{(\prime,*)}{D}^{(*)}$ process. And then, we adopt both the Breit approximation and the non-relativistic normalization to extract the effective potential in the momentum space $\mathcal{V}_E^{\Xi_{c}^{(\prime, *)}{D}^{(*)}\to \Xi_{c}^{(\prime, *)}{D}^{(*)}}(\bm{q})$ using the following equation \cite{Berestetskii:1982qgu}
\begin{eqnarray}
\mathcal{V}_E^{\Xi_{c}^{(\prime, *)}{D}^{(*)}\to \Xi_{c}^{(\prime, *)}{D}^{(*)}}(\bm{q})=-\frac{\mathcal{M}^{\Xi_{c}^{(\prime, *)}{D}^{(*)}\to \Xi_{c}^{(\prime, *)}{D}^{(*)}}(\bm{q})} {\sqrt{2m_{\Xi_{c}^{(\prime, *)}}2m_{{D}^{(*)}}2m_{\Xi_{c}^{(\prime, *)}}2m_{{D}^{(*)}}}}.
\end{eqnarray}
After utilizing the Fourier transformation, the effective potential in the coordinate space $\mathcal{V}_E^{\Xi_{c}^{(\prime, *)}{D}^{(*)}\to \Xi_{c}^{(\prime, *)}{D}^{(*)}}(\bm{r})$ can be calculated \cite{Chen:2016qju, Liu:2019zoy}, i.e.,
\begin{eqnarray}
&&\mathcal{V}_E^{\Xi_{c}^{(\prime, *)}{D}^{(*)}\to \Xi_{c}^{(\prime, *)}{D}^{(*)}}(\bm{r})\nonumber\\ &&=\int\frac{d^3\bm{q}}{(2\pi)^3}e^{i\bm{q}\cdot\bm{r}}\mathcal{V}_E^{\Xi_{c}^{(\prime, *)}{D}^{(*)}\to \Xi_{c}^{(\prime, *)}{D}^{(*)}}(\bm{q})\mathcal{F}^2(q^2,m_{\mathcal{E}}^2).
\end{eqnarray}
Here, we have included the form factor $\mathcal{F}(q^2,m_{\mathcal{E}}^2)$ at two interaction vertices in the Feynman diagram to account for the inner structures of the focused hadrons and the off-shell nature of the exchanged light mesons. Similarly to reproducing  the bound state properties of the deuteron \cite{Machleidt:1987hj,Epelbaum:2008ga,Esposito:2014rxa,Chen:2016qju,Tornqvist:1993ng,Tornqvist:1993vu,Wang:2019nwt,Chen:2017jjn}, we select the monopole-type form factor $\mathcal{F}(q^2,m_{\mathcal{E}}^2) = {(\Lambda^2-m_{\mathcal{E}}^2)}/{(\Lambda^2-q^2)}$ in this work. Here, $\Lambda$ represents the cutoff parameter, the four-momentum of the exchanged light meson is $q$, and the mass of the exchanged light meson is  $m_{\mathcal{E}}$.

For the $AB \to CD$ process if $m_A+m_B>m_C+m_D$, we have $E_A+E_B=\sqrt{m_C^2+\bm k^2}+\sqrt{m_D^2+\bm k^2}$ according to the law of conservation of energy. Thus, the value of $\bm k^2$ can be deduced. From this, we obtain the equations $E_C=\frac{(E_A + E_B)^2 + m_C^2 - m_D^2}{2 (E_A + E_B)}$ and $E_D=\frac{(E_A + E_B)^2 - m_C^2 + m_D^2}{2 (E_A + E_B)}$. For the four-momentum of the exchanged light meson $q$ in the form factor, we have $q^2=q_0^2-\bm q^2$ and $q_0=|E_C-E_A|=|E_D-E_B|$. Thus, $q_0=\frac{\left|m_A^2+m_D^2-m_B^2-m_C^2\right|}{2(E_A+E_B)}$ can be obtained. Using the non-relativistic approximation, we get $q_0=\frac{\left|m_A^2+m_D^2-m_B^2-m_C^2\right|}{2(m_A+m_B)}$. Similarly, $q_0=\frac{\left|m_A^2+m_D^2-m_B^2-m_C^2\right|}{2(m_C+m_D)}$ can be obtained for the $AB \to CD$ process if $m_A+m_B<m_C+m_D$. In particular, there exist $q_0=0$ and $q^2=-\bm q^2$ for the $AB \to AB$ process.

The flavour wave functions $|I,m_{I}\rangle$ for the $\Xi_c^{(\prime,*)}{D}^{(*)}$ systems are constructed as
\begin{eqnarray*}
|1,1\rangle&=&\left|\Xi_c^{(\prime,*)+}{D}^{(*)+}\right\rangle,\nonumber\\
|1,0\rangle&=&-\sqrt{\frac{1}{2}}\left|\Xi_c^{(\prime,*)+}{D}^{(*)0}\right\rangle+\sqrt{\frac{1}{2}}\left|\Xi_c^{(\prime,*)0}{D}^{(*)+}\right\rangle,\nonumber\\
|1,-1\rangle&=&-\left|\Xi_c^{(\prime,*)0}{D}^{(*)0}\right\rangle,\nonumber\\
|0,0\rangle&=&-\sqrt{\frac{1}{2}}\left|\Xi_c^{(\prime,*)+}{D}^{(*)0}\right\rangle-\sqrt{\frac{1}{2}}\left|\Xi_c^{(\prime,*)0}{D}^{(*)+}\right\rangle.
\end{eqnarray*}
Here, the isospins and the corresponding third components of the $\Xi_c^{(\prime,*)}{D}^{(*)}$ systems are denoted as the symbols $I$ and $m_{I}$, respectively. And then, we provide the information on the spin-orbital wave functions $|{}^{2S+1}L_{J}\rangle$ for the $\Xi_c^{(\prime,*)}{D}^{(*)}$ systems as follows
\begin{eqnarray*}
\Xi_c^{(\prime)}{D}&:&\left|{}^{2S+1}L_{J}\right\rangle=\sum_{m,m_L}C^{JM}_{\frac{1}{2}m,Lm_L}\chi_{\frac{1}{2}m}Y_{L m_L},\nonumber\\
\Xi_c^{*}{D}&:&\left|{}^{2S+1}L_{J}\right\rangle=\sum_{m,m_L}C^{JM}_{\frac{3}{2}m,Lm_L}\Phi_{\frac{3}{2}m}Y_{L m_L},\nonumber\\
\Xi_c^{(\prime)}{D}^*&:&\left|{}^{2S+1}L_{J}\right\rangle=\sum_{m,m',m_S,m_L}C^{Sm_S}_{\frac{1}{2}m,1m'}C^{JM}_{Sm_S,Lm_L}\chi_{\frac{1}{2}m}\epsilon_{m'}^{\mu}Y_{L m_L},\nonumber\\
\Xi_c^{*}{D}^*&:&\left|{}^{2S+1}L_{J}\right\rangle=\sum_{m,m',m_S,m_L}C^{Sm_S}_{\frac{3}{2}m,1m'}C^{JM}_{Sm_S,Lm_L}\Phi_{\frac{3}{2}m}^{\mu}\epsilon_{m'}^{\nu}Y_{L m_L}.\nonumber\\
\end{eqnarray*}
Here, the spin, the orbital angular momentum, and the total angular momentum for the $\Xi_c^{(\prime,*)}{D}^{(*)}$ systems are denoted by the symbols $S$, $L$, and $J$, respectively. Furthermore, the spherical harmonic function is denoted by $Y_{L m_L}$.

Given the unique impacts of the $S$-$D$ wave mixing effect and the coupled channel effect in the discussion of the $\Xi_c^{(\prime)}\bar D^{(*)}$ molecular states \cite{Wang:2022mxy}, it is imperative to consider the contribution of the $S$-$D$ wave mixing effect and the coupled channel effect when studying the bound state properties of the $\Xi_c^{(\prime,*)} D^{(*)}$ systems. The $S$-wave and $D$-wave channels $|{}^{2S+1}L_{J}\rangle$ for the $\Xi_c^{(\prime,*)} D^{(*)}$ systems are detailed below
\begin{itemize}
\item  $\Xi_{c}^{(\prime)}{D}$ system:
\begin{eqnarray*}
J^P=\frac{1}{2}^-:\left|{}^2\mathbb{S}_{\frac{1}{2}}\right\rangle,
\end{eqnarray*}
\item  $\Xi_{c}^{*}{D}$ system:
\begin{eqnarray*}
J^P=\frac{3}{2}^-:\left|{}^4\mathbb{S}_{\frac{3}{2}}\right\rangle,\,\left|{}^4\mathbb{D}_{\frac{3}{2}}\right\rangle,
\end{eqnarray*}
\item  $\Xi_{c}^{(\prime)}{D^*}$ system:
\begin{eqnarray*}
    J^P=\frac{1}{2}^-&:&\left|{}^2\mathbb{S}_{\frac{1}{2}}\right\rangle,\,\left|{}^4\mathbb{D}_{\frac{1}{2}}\right\rangle,\nonumber\\
    J^P=\frac{3}{2}^-&:&\left|{}^4\mathbb{S}_{\frac{3}{2}}\right\rangle,\,\left|{}^2\mathbb{D}_{\frac{3}{2}}\right\rangle,\,\left|{}^4\mathbb{D}_{\frac{3}{2}}\right\rangle,
\end{eqnarray*}
\item  $\Xi_{c}^{*}{D^*}$ system:
\begin{eqnarray*}
    J^P=\frac{1}{2}^-&:&\left|{}^2\mathbb{S}_{\frac{1}{2}}\right\rangle,\,\left|{}^4\mathbb{D}_{\frac{1}{2}}\right\rangle,\,\left|{}^6\mathbb{D}_{\frac{1}{2}}\right\rangle,\nonumber\\
    J^P=\frac{3}{2}^-&:&\left|{}^4\mathbb{S}_{\frac{3}{2}}\right\rangle,\,\left|{}^2\mathbb{D}_{\frac{3}{2}}\right\rangle,\,\left|{}^4\mathbb{D}_{\frac{3}{2}}\right\rangle,\,\left|{}^6\mathbb{D}_{\frac{3}{2}}\right\rangle,\nonumber\\
    J^P=\frac{5}{2}^-&:&\left|{}^6\mathbb{S}_{\frac{5}{2}}\right\rangle,\,\left|{}^2\mathbb{D}_{\frac{5}{2}}\right\rangle,\,\left|{}^4\mathbb{D}_{\frac{5}{2}}\right\rangle,\,\left|{}^6\mathbb{D}_{\frac{5}{2}}\right\rangle.
\end{eqnarray*}
\end{itemize}
Here, the spin $S$, the orbital angular momentum $L$, and the total angular momentum $J$ of the relevant channels are denoted by the notation $|^{2S+1}L_J\rangle$.

According to the effective potentials in the coordinate space for the $\Xi_c^{(\prime,*)} D^{(*)}$ systems obtained through the OBE model \cite{Wang:2022mxy,Wang:2020dya,Wang:2019nwt,Wang:2019aoc,Wang:2020bjt,Wang:2021hql,Chen:2018pzd,Yang:2021sue,Wang:2021ajy,Wang:2021yld,Wang:2021aql,Wang:2023aob}, the coupled channel Schr\"{o}dinger equation enables us to discuss the binding energies $E$ and the corresponding spatial wave functions $\phi(r)$ for the $\Xi_c^{(\prime,*)} D^{(*)}$ systems. In addition, the aforementioned spatial wave functions $\phi(r)$ can be used to compute the root-mean-square radius $r_{\rm RMS}$ and the individual channel probabilities $P_i$. These
obtained results can offer the valuable guidance in establishing the mass spectra and investigating the properties of the $\Xi_c^{(\prime,*)} D^{(*)}$-type double-charm molecular pentaquark candidates with single strangeness. In the realistic calculations, we search for the loosely bound state solutions for the $\Xi_c^{(\prime,*)} D^{(*)}$ systems by varying the cutoff parameter within the range of 0.8 to 2.0 GeV, and the reasonable input for discussing the hadronic molecular states is the cutoff parameter about 1.0 GeV in the monopole-type form factor, which is based on the previous experiences of reproducing the deuteron's bound state properties within the OBE model \cite{Machleidt:1987hj,Epelbaum:2008ga,Esposito:2014rxa,Chen:2016qju,Tornqvist:1993ng,Tornqvist:1993vu,Wang:2019nwt,Chen:2017jjn}. Considering that the hadronic molecular state is the loosely bound state comprising the hadronic states, the small binding energy and the large root-mean-square radius are the important characteristics of the most promising hadronic molecular candidate to avoid the excessive overlap between the corresponding  constituent hadrons in their spatial distribution \cite{Chen:2016qju,Chen:2017xat}.

\subsection{The mass spectra and the corresponding spatial wave functions of the $\Xi_c^{(\prime,*)} D^{(*)}$ molecular pentaquarks}

\begin{figure}[htbp]
\centering
  \includegraphics[width=0.49\textwidth]{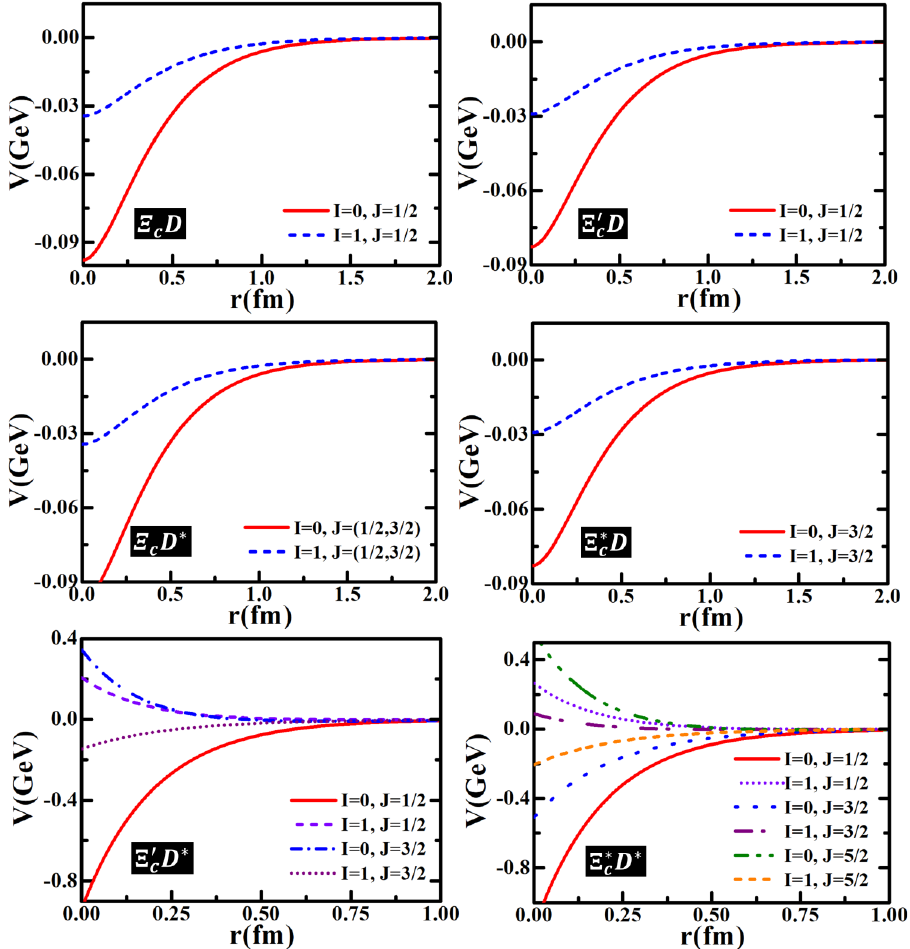}
  \caption{The interaction potentials of the $\Xi_c^{(\prime,*)} D^{(*)}$ systems, where the cutoff value $\Lambda$ is fixed as 1.00 GeV.}\label{potentialS}
\end{figure}

In Fig. \ref{potentialS}, we present the interaction potentials of the $\Xi_c^{(\prime,*)} D^{(*)}$ systems, where the cutoff value $\Lambda$ is fixed as 1.00 GeV.  From Fig. \ref{potentialS}, we can see which states are more likely to form the loosely bound states. For example, the $\Xi_c D$ state with $I(J^P)=0(1/2^-)$ is more likely to generate a loosely bound state compared to the $\Xi_c D$ state with $I(J^P)=1(1/2^-)$, which is mainly due to the fact that the $\Xi_c D$ state with $I(J^P)=0(1/2^-)$ has stronger attractive interactions.

\subsubsection{The $S$-wave $\Xi_c D$ system}

In Table \ref{massspectra1}, we list the bound state properties for the $\Xi_c D$ state with $I(J^P)=0(1/2^-)$ revealed by both the single channel analysis and the coupled channel analysis.

\renewcommand\tabcolsep{0.32cm}
\renewcommand{\arraystretch}{1.50}
\begin{table}[!htbp]
\caption{The obtained bound state solutions for the $\Xi_c D$ state with $I(J^P)=0(1/2^-)$. The cutoff parameter $\Lambda$, binding energy $E$, and root-mean-square radius $r_{\rm RMS}$ are measured in units of $\rm{GeV}$, $\rm {MeV}$, and $\rm {fm}$, respectively. It is worth mentioning that the primary probability of the corresponding channels is emphasized in bold font.}\label{massspectra1}
\begin{tabular}{cccc}\toprule[1pt]\toprule[1pt]
\multicolumn{4}{c}{Single channel analysis}\\\hline
$\Lambda$ &$E$  &$r_{\rm RMS}$\\
$1.18$&$-0.28$&$4.97$\\
$1.28$&$-4.80$&$1.67$\\
$1.37$&$-12.85$&$1.12$\\\hline
\multicolumn{4}{c}{Coupled channel analysis}\\\hline
$\Lambda$ &$E$  &$r_{\rm RMS}$& P($\Xi_c D/\Xi_c^{\prime} D/\Xi_c D^*/\Xi_c^{\prime} D^*/\Xi_c^{*} D^*$)\\
1.01&$-0.27$ &4.93&\textbf{98.12}/0.02/0.04/0.97/0.86\\
1.04&$-3.24$ &1.84&\textbf{90.96}/0.26/0.61/5.30/2.86\\
1.07&$-13.36$ &0.81&\textbf{53.79}/4.49/13.08/27.63/1.00\\
\bottomrule[1pt]\bottomrule[1pt]
\end{tabular}
\end{table}

By performing the single channel analysis, the isoscalar $\Xi_c D$ state with $J^P=1/2^-$ displays the loosely bound state solutions when the cutoff value $\Lambda$ exceeds 1.18 GeV. Furthermore, the bound state properties for the $\Xi_c D$ state with $I(J^P)=0(1/2^-)$ can also be explored by taking into account the coupled channel effect involving the $\Xi_c D$, $\Xi_c^{\prime} D$, $\Xi_c D^*$, $\Xi_c^{\prime} D^*$, and $\Xi_c^{*} D^*$ channels. By numerically solving the coupled channel Schr\"{o}dinger equation, the loosely bound state solutions can be obtained for the $\Xi_c D$ state with $I(J^P)=0(1/2^-)$ when the cutoff value $\Lambda$ is greater than 1.01 GeV, where the primary contributor is the $\Xi_c D$ channel. As the cutoff parameter increases, the binding energy of the $\Xi_c D$ state with $I(J^P)=0(1/2^-)$ increases, and other channels such as the $\Xi_cD^*$ and $\Xi_c^{\prime}D^*$ become increasingly significant. In conclusion, the $\Xi_c D$ state with $I(J^P)=0(1/2^-)$ can be regarded as the ideal candidate of the double-charm molecular pentaquark with single strangeness, which can be deemed as the partner of the $\Xi_c \bar D$ molecular state with $I(J^P)=0(1/2^-)$ \cite{Wang:2022mxy}. Subsequent experimental exploration of the $\Xi_c D$ molecular state with $I(J^P)=0(1/2^-)$ can potentially clarify the explanation of the observed $P_{\psi s}^{\Lambda}(4338)$ state \cite{LHCb:2022ogu} as the $\Xi_c\bar D$ molecular state with $I(J^P)=0(1/2^-)$ \cite{Chen:2016ryt,Wu:2010vk,Hofmann:2005sw,Anisovich:2015zqa,Feijoo:2015kts,Lu:2016roh,Xiao:2019gjd,Chen:2015sxa,Wang:2019nvm,Weng:2019ynv}.

For the $\Xi_c D$ state with $I(J^P)=1(1/2^-)$, our analysis indicates that  it fails to exhibit the loosely bound state solutions when varying the cutoff parameter between 0.8 to 2.0 GeV even after increasing the contribution of the coupled channel effect. Consequently, it is not recommended to consider the $\Xi_c D$ state with $I(J^P)=1(1/2^-)$ as the candidate of the double-charm molecular pentaquark with single strangeness.

\subsubsection{The $S$-wave $\Xi_c^{\prime} D$ system}

In Table \ref{massspectra2}, we present the bound state properties for the $\Xi_c^{\prime} D$ state with $I(J^P)=0(1/2^-)$, where we conducted  both the single channel analysis and the coupled channel analysis.

\renewcommand\tabcolsep{0.40cm}
\renewcommand{\arraystretch}{1.50}
\begin{table}[!htbp]
\caption{The obtained bound state solutions for the $\Xi_c^{\prime} D$ state with $I(J^P)=0(1/2^-)$. The cutoff parameter $\Lambda$, binding energy $E$, and root-mean-square radius $r_{\rm RMS}$ are measured in units of $\rm{GeV}$, $\rm {MeV}$, and $\rm {fm}$, respectively. It is worth noting that the primary probability of the corresponding channels is highlighted in bold font.}\label{massspectra2}
\begin{tabular}{cccc}\toprule[1pt]\toprule[1pt]
\multicolumn{4}{c}{Single channel analysis}\\\hline
$\Lambda$ &$E$  &$r_{\rm RMS}$\\
$1.24$&$-0.25$&$5.10$\\
$1.36$&$-4.72$&$1.66$\\
$1.47$&$-12.81$&$1.10$\\\hline
\multicolumn{4}{c}{Coupled channel analysis}\\\hline
$\Lambda$ &$E$  &$r_{\rm RMS}$& P($\Xi_c^{\prime} D/\Xi_c D^*/\Xi_c^{\prime} D^*/\Xi_c^{*} D^*$)\\
0.89&$-0.61$ &3.61&\textbf{86.51}/9.61/3.09/0.78\\
0.92&$-6.26$ &1.19&\textbf{62.84}/25.50/9.34/2.32\\
0.94&$-12.83$ &0.86&\textbf{51.90}/31.89/13.01/3.20\\
\bottomrule[1pt]\bottomrule[1pt]
\end{tabular}
\end{table}

Within the single channel analysis, the $\Xi_c^{\prime} D$ state with $I(J^P)=0(1/2^-)$ exhibits the loosely bound state solutions if the cutoff parameter exceeds 1.24 GeV. Accounting for the coupled channel effect and considering the cutoff parameter greater than 0.89 GeV, the $\Xi_c^{\prime} D$ state with $I(J^P)=0(1/2^-)$ exists the loosely bound state solutions. The $\Xi_c^{\prime} D$ channel contributes significantly, and the $\Xi_cD^*$ and $\Xi_c^{\prime}D^*$ channels also play the important roles when the $\Xi_c^{\prime} D$ state with $I(J^P)=0(1/2^-)$ exhibits the larger bound energy. Consequently, we propose the $\Xi_c^{\prime} D$ state with $I(J^P)=0(1/2^-)$ as the promising candidate of the double-charm molecular pentaquark with single strangeness.

When the cutoff parameter ranges from 0.8 to 2.0 GeV, the $\Xi_c^{\prime} D$ state with $I(J^P)=1(1/2^-)$ does not emerge the loosely bound state solutions, even if the coupled channel effect is taken into account. Thus, we cannot recommend the $\Xi_c^{\prime} D$ state with $I(J^P)=1(1/2^-)$ as the candidate of the double-charm molecular pentaquark with single strangeness.

\subsubsection{The $S$-wave $\Xi_c D^{*}$ system}

It is well known that the $P_{\psi s}^{\Lambda}(4459)$ state \cite{LHCb:2020jpq} can be associated with the $\Xi_c \bar D^{*}$ molecular candidate \cite{Chen:2016ryt,Wu:2010vk,Hofmann:2005sw,Anisovich:2015zqa,Feijoo:2015kts,Lu:2016roh,Xiao:2019gjd,Chen:2015sxa,Wang:2019nvm,Weng:2019ynv}, the study of the $\Xi_c D^{*}$ molecular states can enhance the understanding of the inner structure of the $P_{cs}(4459)$ \cite{Karliner:2022erb,Wang:2022mxy}. In Table \ref{massspectra3}, the bound state solutions for the $\Xi_c D^{*}$ system are summarised.

\renewcommand\tabcolsep{0.25cm}
\renewcommand{\arraystretch}{1.50}
\begin{table}[!htbp]
\caption{The obtained bound state solutions for the $\Xi_c D^{*}$ system. The cutoff parameter $\Lambda$, binding energy $E$, and root-mean-square radius $r_{\rm RMS}$ are measured in units of $\rm{GeV}$, $\rm {MeV}$, and $\rm {fm}$, respectively. Notably, the primary probability of the corresponding channels is emphasized in bold font.}\label{massspectra3}
\begin{tabular}{c|cccc}\toprule[1pt]\toprule[1pt]
\multicolumn{1}{c|}{$I(J^P)$}&\multicolumn{4}{c}{Bound state solutions}\\\midrule[1.0pt]
\multirow{15}{*}{$0(\frac{1}{2}^{-})$}&\multicolumn{4}{c}{Single channel analysis}\\
\cline{2-5}
&$\Lambda$ &$E$  &$r_{\rm RMS}$\\
&$1.17$&$-0.38$&$4.57$\\
&$1.26$&$-4.57$&$1.68$\\
&$1.35$&$-12.71$&$1.11$\\
\cline{2-5}
&\multicolumn{4}{c}{$S$-$D$ wave mixing analysis}\\
\cline{2-5}
&$\Lambda$ &$E$  &$r_{\rm RMS}$& P(${}^2\mathbb{S}_{\frac{1}{2}}/{}^4\mathbb{D}_{\frac{1}{2}}$)\\
&$1.17$&$-0.38$&$4.57$&\textbf{100.00}/$\mathcal{O}(0)$\\
&$1.26$&$-4.57$&$1.68$&\textbf{100.00}/$\mathcal{O}(0)$\\
&$1.35$&$-12.71$&$1.11$&\textbf{100.00}/$\mathcal{O}(0)$\\
\cline{2-5}
&\multicolumn{4}{c}{Coupled channel analysis}\\
\cline{2-5}
&$\Lambda$ &$E$  &$r_{\rm RMS}$& P($\Xi_c D^*/\Xi_c^{\prime} D^*/\Xi_c^{*} D^*$)\\
&0.94&$-0.30$ &4.70&\textbf{96.67}/2.79/0.54\\
&0.96&$-3.92$ &1.61&\textbf{87.54}/10.56/1.90\\
&0.98&$-11.92$ &0.96&\textbf{78.00}/18.79/3.21\\\midrule[1.0pt]
\multirow{15}{*}{$0(\frac{3}{2}^{-})$}&\multicolumn{4}{c}{Single channel analysis}\\
\cline{2-5}
&$\Lambda$ &$E$  &$r_{\rm RMS}$\\
&$1.17$&$-0.38$&$4.57$\\
&$1.26$&$-4.57$&$1.68$\\
&$1.35$&$-12.71$&$1.11$\\
\cline{2-5}
&\multicolumn{4}{c}{$S$-$D$ wave mixing analysis}\\
\cline{2-5}
&$\Lambda$ &$E$  &$r_{\rm RMS}$& P(${}^4\mathbb{S}_{\frac{3}{2}}/{}^2\mathbb{D}_{\frac{3}{2}}/{}^4\mathbb{D}_{\frac{3}{2}}$)\\
&$1.17$&$-0.38$&$4.57$&\textbf{100.00}/$\mathcal{O}(0)$/$\mathcal{O}(0)$\\
&$1.26$&$-4.57$&$1.68$&\textbf{100.00}/$\mathcal{O}(0)$/$\mathcal{O}(0)$\\
&$1.35$&$-12.71$&$1.11$&\textbf{100.00}/$\mathcal{O}(0)$/$\mathcal{O}(0)$\\
\cline{2-5}
&\multicolumn{4}{c}{Coupled channel analysis}\\
\cline{2-5}
&$\Lambda$ &$E$  &$r_{\rm RMS}$& P($\Xi_c D^*/\Xi_c^* D/\Xi_c^{\prime} D^*/\Xi_c^{*} D^*$)\\
&0.88&$-0.33$ &4.57&\textbf{93.92}/4.19/0.18/1.72\\
&0.91&$-4.47$ &1.50&\textbf{80.82}/12.14/0.73/6.31\\
&0.94&$-13.23$ &0.93&\textbf{71.61}/15.86/1.45/11.08\\
\bottomrule[1pt]\bottomrule[1pt]
\end{tabular}
\end{table}

For the single channel analysis, the $\Xi_c D^{*}$ states with $I(J^P)=0(1/2^-,3/2^-)$ can form the loosely bound states when the cutoff parameter is greater than 1.17 GeV. If the cutoff parameters are identical, the $\Xi_c D^{*}$ states with $I(J^P)=0(1/2^-,3/2^-)$ exhibit the comparable binding properties, which can be attributed to the same interaction described by the OBE model for both states in the single channel analysis. For the $\Xi_c D^{*}$ states with $I(J^P)=0(1/2^-,3/2^-)$, we can also analyze the impact of the $S$-$D$ wave mixing effect for their bound state properties. However, the interactions of the $\Xi_c D^{*}$ system do not exist the tensor interactions. Thus, the $S$-$D$ wave mixing effect does not influence the bound state properties for the $\Xi_c D^{*}$ states with $I(J^P)=0(1/2^-,3/2^-)$, where the contribution of the corresponding $D$-wave channels is zero. In conclusion, the $\Xi_c D^{*}$ states with $I(J^P)=0(1/2^-,3/2^-)$ can be considered as the ideal candidates of the double-charm molecular pentaquarks with single strangeness, and their binding properties are identical when the corresponding cutoff parameters are equal following the single channel analysis and the $S$-$D$ wave mixing analysis. Here, we need to mention that there exists the similar binding properties of the $\Xi_c \bar D^{*}$ molecular states with $I(J^P)=0(1/2^-,3/2^-)$ \cite{Wang:2022mxy}.

In order to investigate the bound state properties of the $\Xi_c D^{*}$ states with $I(J^P)=0(1/2^-,3/2^-)$, it is essential to take into account the influence of the coupled channel effect.  Regarding the $\Xi_c D^{*}$ state with $I(J^P)=0(1/2^-)$, the coupling between the $\Xi_c D^*$, $\Xi_c^{\prime} D^*$, and $\Xi_c^{*} D^*$ channels can be taken into account, and the loosely bound state solutions arise when the cutoff parameter is set to 0.94 GeV, where the main contribution is provided by the $\Xi_c D^*$ channel, with the $\Xi_c^{\prime} D^*$ and $\Xi_c^{*} D^*$ channels also playing the role. For the $\Xi_c D^{*}$ state with $I(J^P)=0(3/2^-)$, the coupling of the $\Xi_c D^*$, $\Xi_c^* D$, $\Xi_c^{\prime} D^*$, and $\Xi_c^{*} D^*$ channels may affect its binding properties. When setting the cutoff parameter to 0.88 GeV, the $\Xi_c D^{*}$ state with $I(J^P)=0(3/2^-)$ begins to appear the loosely bound state solutions, and the $\Xi_c D^*$ channel is the principal contributor, with the $\Xi_c^* D$ and $\Xi_c^{*} D^*$ channels playing a more significant role as the binding energy increases. By introducing the coupled channel effect, the $\Xi_c D^{*}$ states with $I(J^P)=0(1/2^-,3/2^-)$ exhibit different binding properties when both states take the same cutoff parameter, which is analogous to the bound state properties of the $\Xi_c \bar D^{*}$ molecular states with $I(J^P)=0(1/2^-,3/2^-)$ \cite{Wang:2022mxy}. Therefore, this proposal suggests that the upcoming experiments should investigate the $\Xi_c D^{*}$ molecular states with $I(J^P)=0(1/2^-,3/2^-)$, which can provide the crucial check for the double peak structures located just below the threshold of the $\Xi_c\bar D^*$ channel for the $P_{\psi s}^\Lambda(4459)$ state \cite{Karliner:2022erb,Wang:2022mxy}.

In addition to discussing the bound state properties of the $\Xi_c D^{*}$ states with $I(J^P)=0(1/2^-,3/2^-)$, we also examine the bound state properties of the $\Xi_c D^{*}$ states with $I(J^P)=1(1/2^-,3/2^-)$. Unfortunately, the cutoff parameter varies between 0.8 to 2.0 GeV, there is no evidence of the existence of the bound state solutions for the $\Xi_c D^{*}$ states with $I(J^P)=1(1/2^-,3/2^-)$ despite the increase in the $S$-$D$ wave mixing effect and the coupled channel effect. Hence, the $\Xi_c D^{*}$ states with $I(J^P)=1(1/2^-,3/2^-)$ cannot be deemed as the candidates of the double-charm molecular pentaquarks with single strangeness.

\subsubsection{The $S$-wave $\Xi_c^{*} D$ system}

In Table \ref{massspectra4}, we present the bound state properties for the $\Xi_c^{*} D$ state with $I(J^P)=0(3/2^-)$. Here, we need to mention that there are no loosely bound state solutions for the $\Xi_c^{*} D$ state with $I(J^P)=1(3/2^-)$ when the cutoff parameter variations ranging between 0.8 to 2.0 GeV, despite an increase in the $S$-$D$ wave mixing effect and the coupled channel effect. As a result, this rules out the $\Xi_c^{*} D$ state with $I(J^P)=1(3/2^-)$ as the candidate of the double-charm molecular pentaquark with single strangeness.

\renewcommand\tabcolsep{0.50cm}
\renewcommand{\arraystretch}{1.50}
\begin{table}[!htbp]
\caption{The obtained bound state solutions for the $\Xi_c^{*} D$ state with $I(J^P)=0(3/2^-)$. The cutoff parameter $\Lambda$, binding energy $E$, and root-mean-square radius $r_{\rm RMS}$ are measured in units of $\rm{GeV}$, $\rm {MeV}$, and $\rm {fm}$, respectively. It should be emphasized that the primary probability of the corresponding channels is highlighted in bold font.}\label{massspectra4}
\begin{tabular}{cccc}\toprule[1pt]\toprule[1pt]
\multicolumn{4}{c}{Single channel analysis}\\\hline
$\Lambda$ &$E$  &$r_{\rm RMS}$\\
$1.24$&$-0.33$&$4.78$\\
$1.35$&$-4.45$&$1.70$\\
$1.46$&$-12.48$&$1.11$\\\hline
\multicolumn{4}{c}{$S$-$D$ wave mixing analysis}\\\hline
$\Lambda$ &$E$  &$r_{\rm RMS}$& P(${}^4\mathbb{S}_{\frac{3}{2}}/{}^4\mathbb{D}_{\frac{3}{2}}$)\\
$1.24$&$-0.33$&$4.78$&\textbf{100.00}/$\mathcal{O}(0)$\\
$1.35$&$-4.45$&$1.70$&\textbf{100.00}/$\mathcal{O}(0)$\\
$1.46$&$-12.48$&$1.11$&\textbf{100.00}/$\mathcal{O}(0)$\\\hline
\multicolumn{4}{c}{Coupled channel analysis}\\\hline
$\Lambda$ &$E$  &$r_{\rm RMS}$& P($\Xi_c^* D/\Xi_c^{\prime} D^*/\Xi_c^{*} D^*$)\\
1.17&$-0.25$ &5.04&\textbf{98.84}/0.04/1.12\\
1.21&$-1.54$ &2.62&\textbf{94.82}/0.45/4.73\\
1.25&$-7.70$ &1.01&\textbf{51.20}/9.44/39.36\\
\bottomrule[1pt]\bottomrule[1pt]
\end{tabular}
\end{table}

When the cutoff parameter is set to 1.24 GeV during the single channel analysis, the $\Xi_c^{*} D$ state with $I(J^P)=0(3/2^-)$ emerges the loosely bound state solutions. Similarly to the $\Xi_c D$ and $\Xi_c^{\prime} D$ systems, the interaction of the $\Xi_c^{*} D$ system lacks the tensor interaction. As a result, both the single channel analysis and the $S$-$D$ wave mixing analysis can give the same bound state solutions for the $\Xi_c^{*} D$ state with $I(J^P)=0(3/2^-)$ when the cutoff parameter takes the same value, and the contribution from the $D$-wave channel is absent. Furthermore, the coupled channel effect including the $\Xi_c^* D$, $\Xi_c^{\prime} D^*$, and $\Xi_c^{*} D^*$ channels also affects the bound state properties of the $\Xi_c^{*} D$ state with $I(J^P)=0(3/2^-)$. When the cutoff parameter is 1.17 GeV, the $\Xi_c^{*} D$ state with $I(J^P)=0(3/2^-)$ appears the loosely bound state solutions. When the binding energy is shallow, the $\Xi_c^* D$ channel dominates, but the contribution of the $\Xi_c^{*} D^*$ channel increases gradually with the binding energy. Consequently, the $\Xi_c^{*} D$ state with $I(J^P)=0(3/2^-)$ is suggested as the preferred candidate of the double-charm molecular pentaquark with single strangeness.

\subsubsection{The $S$-wave $\Xi_c^{\prime} D^{*}$ system}

In Table \ref{massspectra5}, the bound state properties of the $\Xi_c^{\prime} D^{*}$ system are presented, where the influences of the $S$-$D$ wave mixing effect and the coupled channel effect are analyzed. When varying the cutoff parameter between 0.8 to 2.0 GeV, the $\Xi_c^{\prime} D^{*}$ states with $I(J^P)=1(1/2^-,3/2^-)$ fail to generate the loosely bound state solutions, even though the realistic calculations have taken into account several effects like the $S$-$D$ wave mixing effect and the coupled channel effect. Thus, the $\Xi_c^{\prime} D^{*}$ states with $I(J^P)=1(1/2^-,3/2^-)$ are not supported as the candidates of the double-charm molecular pentaquarks with single strangeness.

\renewcommand\tabcolsep{0.36cm}
\renewcommand{\arraystretch}{1.50}
\begin{table}[!htbp]
\caption{The obtained bound state solutions for the $\Xi_c^{\prime} D^{*}$ system. The cutoff parameter $\Lambda$, binding energy $E$, and root-mean-square radius $r_{\rm RMS}$ are expressed in units of $\rm{GeV}$, $\rm {MeV}$, and $\rm {fm}$, respectively. Emphasis is added in bold to highlight the primary probability of the corresponding channels.}\label{massspectra5}
\begin{tabular}{c|cccc}\toprule[1pt]\toprule[1pt]
\multicolumn{1}{c|}{$I(J^P)$}&\multicolumn{4}{c}{Bound state solutions}\\\midrule[1.0pt]
\multirow{15}{*}{$0(\frac{1}{2}^{-})$}&\multicolumn{4}{c}{Single channel analysis}\\
\cline{2-5}
&$\Lambda$ &$E$  &$r_{\rm RMS}$\\
&$0.92$&$-0.66$&$3.62$\\
&$0.96$&$-4.94$&$1.52$\\
&$1.00$&$-13.74$&$0.99$\\
\cline{2-5}
&\multicolumn{4}{c}{$S$-$D$ wave mixing analysis}\\
\cline{2-5}
&$\Lambda$ &$E$  &$r_{\rm RMS}$& P(${}^2\mathbb{S}_{\frac{1}{2}}/{}^4\mathbb{D}_{\frac{1}{2}}$)\\
&$0.91$&$-0.53$&$3.96$&\textbf{99.67}/0.33\\
&$0.95$&$-4.21$&$1.64$&\textbf{99.55}/0.45\\
&$0.99$&$-12.03$&$1.06$&\textbf{99.60}/0.40\\
\cline{2-5}
&\multicolumn{4}{c}{Coupled channel analysis}\\
\cline{2-5}
&$\Lambda$ &$E$  &$r_{\rm RMS}$& P($\Xi_c^{\prime} D^*/\Xi_c^{*} D^*$)\\
&0.90&$-0.63$ &3.67&\textbf{98.85}/1.15\\
&0.93&$-4.15$ &1.60&\textbf{95.91}/4.09\\
&0.96&$-11.63$ &1.02&\textbf{90.62}/9.38\\\midrule[1.0pt]
\multirow{15}{*}{$0(\frac{3}{2}^{-})$}&\multicolumn{4}{c}{Single channel analysis}\\
\cline{2-5}
&$\Lambda$ &$E$  &$r_{\rm RMS}$\\
&$1.57$&$-0.31$&$4.94$\\
&$1.79$&$-2.77$&$2.18$\\
&$2.00$&$-6.69$&$1.53$\\
\cline{2-5}
&\multicolumn{4}{c}{$S$-$D$ wave mixing analysis}\\
\cline{2-5}
&$\Lambda$ &$E$  &$r_{\rm RMS}$& P(${}^4\mathbb{S}_{\frac{3}{2}}/{}^2\mathbb{D}_{\frac{3}{2}}/{}^4\mathbb{D}_{\frac{3}{2}}$)\\
&$1.47$&$-0.30$&$5.00$&\textbf{99.22}/0.13/0.65\\
&$1.74$&$-3.10$&$2.11$&\textbf{98.82}/0.18/1.00\\
&$2.00$&$-7.82$&$1.46$&\textbf{99.05}/0.14/0.81\\
\cline{2-5}
&\multicolumn{4}{c}{Coupled channel analysis}\\
\cline{2-5}
&$\Lambda$ &$E$  &$r_{\rm RMS}$& P($\Xi_c^{\prime} D^*/\Xi_c^{*} D^*$)\\
&1.13&$-1.04$ &2.90&\textbf{84.05}/5.95\\
&1.15&$-6.12$ &1.17&\textbf{60.71}/39.29\\
&1.16&$-10.09$ &0.91&\textbf{52.22}/47.28\\
\bottomrule[1pt]\bottomrule[1pt]
\end{tabular}
\end{table}

When performing the single channel analysis with the cutoff parameter over 0.92 GeV, the $\Xi_c^{\prime} D^{*}$ state with $I(J^P)=0(1/2^-)$ can form the loosely bound state. The bound state properties of the $\Xi_c^{\prime} D^{*}$ state with $I(J^P)=0(1/2^-)$ also can be discussed with the inclusion of the $S$-$D$ wave mixing effect, but the $S$-$D$ wave mixing effect does not have the significant impact on its bound state properties, as the contribution of the $S$-wave channel is over 99\%. Furthermore, we can discuss the bound state properties of the $\Xi_c^{\prime} D^{*}$ state with $I(J^P)=0(1/2^-)$ by incorporating the coupled channel effect, which makes the cutoff parameter smaller when the same binding energy is achieved in the single channel analysis or the $S$-$D$ wave mixing analysis, of which the $\Xi_c^{\prime} D^*$ channel has the notable contribution of over 90 percent. Based on previous discussions, we propose the $\Xi_c^{\prime} D^{*}$ state with $I(J^P)=0(1/2^-)$ as the preferred candidate of the double-charm molecular pentaquark with single strangeness.

For the $\Xi_c^{\prime} D^{*}$ state with $I(J^P)=0(3/2^-)$, when the cutoff parameter is 1.57 GeV in the single channel analysis, it has the loosely bound state solutions. After increasing the $S$-$D$ wave mixing effect, the loosely bound state solutions still exist with the cutoff parameter at 1.47 GeV, where the total contribution of the $D$-wave channels is less than 1 percent. Furthermore, when the contribution of the coupled channel effect by coupling the $\Xi_c^{\prime} D^*$ and $\Xi_c^{*} D^*$ channels is further increased and the cutoff parameter is 1.13 GeV, the $\Xi_c^{\prime} D^{*}$ state with $I(J^P)=0(3/2^-)$ appears the loosely bound state solutions. At low binding energy, the $\Xi_c^{\prime} D^*$ channel dominates initially, but as the binding energy increases, the $\Xi_c^{*} D^*$ channel also becomes a significant contribution.  In conclusion, the $\Xi_c^{\prime} D^{*}$ state with $I(J^P)=0(3/2^-)$ can be recommended as the suitable candidate of the double-charm molecular pentaquark with single strangeness.

\subsubsection{The $S$-wave $\Xi_c^{*} D^{*}$ system}

In Table \ref{massspectra6}, the bound state solutions for the $\Xi_c^{*} D^{*}$ system are displayed. In the following, we analyze the bound state properties for the $\Xi_c^{*} D^{*}$ system.

\renewcommand\tabcolsep{0.07cm}
\renewcommand{\arraystretch}{1.50}
\begin{table}[!htbp]
\caption{The obtained bound state solutions for the $\Xi_c^{*} D^{*}$ system. The cutoff parameter $\Lambda$, binding energy $E$, and root-mean-square radius $r_{\rm RMS}$ are expressed in units of $\rm{GeV}$, $\rm {MeV}$, and $\rm {fm}$, respectively. The primary probability of the corresponding channels is indicated in bold font.}\label{massspectra6}
\begin{tabular}{ccc|cccc}\toprule[1pt]\toprule[1pt]
\multicolumn{3}{c|}{Single channel analysis}&\multicolumn{4}{c}{$S$-$D$ wave mixing analysis}\\\midrule[1.0pt]
\multicolumn{7}{c}{$I(J^P)=0(\frac{1}{2}^{-})$}\\
\cline{1-7}
$\Lambda$ &$E$  &$r_{\rm RMS}$ &$\Lambda$ &$E$  &$r_{\rm RMS}$ &P(${}^2\mathbb{S}_{\frac{1}{2}}/{}^4\mathbb{D}_{\frac{1}{2}}/{}^6\mathbb{D}_{\frac{1}{2}})$\\
$0.87$&$-0.61$&$3.71$&      0.85&$-0.43$ &4.24&\textbf{99.38}/0.37/0.25\\
$0.91$&$-4.67$&$1.54$&      0.90&$-4.71$ &1.57&\textbf{99.10}/0.55/0.35\\
$0.95$&$-13.48$&$1.00$&     0.94&$-12.70$ &1.04&\textbf{99.18}/0.51/0.31\\\midrule[1.0pt]
\multicolumn{7}{c}{$I(J^P)=0(\frac{3}{2}^{-})$}\\
\cline{1-7}
$\Lambda$ &$E$  &$r_{\rm RMS}$ &$\Lambda$ &$E$  &$r_{\rm RMS}$ &P(${}^4\mathbb{S}_{\frac{3}{2}}/{}^2\mathbb{D}_{\frac{3}{2}}/{}^4\mathbb{D}_{\frac{3}{2}}/{}^6\mathbb{D}_{\frac{3}{2}})$\\
1.03&$-0.53$ &3.98&      1.00&$-0.31$ &4.74&\textbf{99.28}/0.20/0.47/0.06\\
1.08&$-4.74$ &1.56&      1.06&$-4.34$ &1.66&\textbf{98.81}/0.33/0.77/0.09\\
1.13&$-13.34$ &1.01&     1.12&$-13.70$ &1.02&\textbf{98.93}/0.30/0.69/0.08\\\midrule[1.0pt]
\multicolumn{7}{c}{$I(J^P)=0(\frac{5}{2}^{-})$}\\
\cline{1-7}
$\Lambda$ &$E$  &$r_{\rm RMS}$ &$\Lambda$ &$E$  &$r_{\rm RMS}$ &P(${}^6\mathbb{S}_{\frac{5}{2}}/{}^2\mathbb{D}_{\frac{5}{2}}/{}^4\mathbb{D}_{\frac{5}{2}}/{}^6\mathbb{D}_{\frac{5}{2}})$\\
1.65&$-0.31$ &4.96&      1.53&$-0.32$ &4.95&\textbf{98.96}/0.07/0.07/0.95\\
1.83&$-2.08$ &2.49&      1.77&$-2.50$ &2.34&\textbf{98.42}/0.09/0.05/1.44\\
2.00&$-4.96$ &1.75&     2.00&$-6.31$ &1.63&\textbf{98.58}/0.08/0.04/1.30\\\midrule[1.0pt]
\multicolumn{7}{c}{$I(J^P)=1(\frac{5}{2}^{-})$}\\
\cline{1-7}
&&&$\Lambda$ &$E$  &$r_{\rm RMS}$ &P(${}^6\mathbb{S}_{\frac{5}{2}}/{}^2\mathbb{D}_{\frac{5}{2}}/{}^4\mathbb{D}_{\frac{5}{2}}/{}^6\mathbb{D}_{\frac{5}{2}})$\\
&&&     1.74&$-0.27$ &4.83&\textbf{99.48}/0.07/0.02/0.44\\
&&&     1.87&$-2.01$ &2.24&\textbf{98.85}/0.15/0.04/0.95\\
&&&     2.00&$-5.50$ &1.42&\textbf{98.32}/0.22/0.06/1.39\\
\bottomrule[1pt]\bottomrule[1pt]
\end{tabular}
\end{table}

When conducting the single channel analysis with the cutoff parameter of 0.87 GeV, the $\Xi_c^{*} D^{*}$ state with $I(J^P)=0(1/2^-)$ has the loosely bound state solutions. After increasing the mixing effect between the $S$ and $D$ waves, the cutoff parameter decreases to 0.85 MeV when the $\Xi_c^{*} D^{*}$ state with $I(J^P)=0(1/2^-)$ appears the loosely bound state solutions, and the $D$-wave channel contribution is less than 1\%. Thus, the $\Xi_c^{*} D^{*}$ state with $I(J^P)=0(1/2^-)$ can be considered as the ideal candidate of the double-charm molecular pentaquark with single strangeness.  When considering only the $S$-wave contribution and the cutoff parameter is greater than 1.03 GeV, the $\Xi_c^{*} D^{*}$ state with $I(J^P)=0(3/2^-)$ can form the loosely bound state. Once the contribution of the $D$-wave channels is increased, the cutoff parameter for the $\Xi_c^{*} D^{*}$ state with $I(J^P)=0(3/2^-)$ to form the loosely bound state decreases to 1.00 GeV, where the total contribution of the $D$-wave channels is less than 2 percent. In conclusion, the $\Xi_c^{*} D^{*}$ state with $I(J^P)=0(3/2^-)$ can be regarded as the preferred  candidate of the double-charm molecular pentaquark with single strangeness. In both the single channel analysis and the $S$-$D$ wave mixing analysis, the $\Xi_c^{*} D^{*}$ state with $I(J^P)=0(5/2^-)$ starts to appear the loosely bound state solutions when the cutoff parameters are taken to be 1.65 GeV and 1.53 GeV, respectively. Here, the $S$-wave channel plays the major role and its contribution is greater than 98 percent. Consequently, we suggest that the $\Xi_c^{*} D^{*}$ state with $I(J^P)=0(5/2^-)$ is the essential candidate of the  double-charm molecular pentaquark with single strangeness.

For the $\Xi_c^{*} D^{*}$ states with $I(J^P)=1(1/2^-,3/2^-)$, the loosely bound state solutions are absent when the cutoff parameter is varied between 0.8 to 2.0 GeV and the $S$-$D$ wave mixing effect is taken into account. Thus, the $\Xi_c^{*} D^{*}$ states with $I(J^P)=1(1/2^-,3/2^-)$ are not supported as the candidates of the double-charm molecular pentaquarks with single strangeness. When conducting the single channel analysis with the cutoff parameter within the range of 0.8 to 2.0 GeV, the $\Xi_c^{*} D^{*}$ state with $I(J^P)=1(5/2^-)$ is unable to form the loosely bound state. Considering the $S$-$D$ wave mixing analysis, the $\Xi_c^{*} D^{*}$ state with $I(J^P)=1(5/2^-)$ can form the loosely bound state when the cutoff parameter is 1.74 GeV. Thus, the $\Xi_c^{*} D^{*}$ state with $I(J^P)=1(5/2^-)$ can be considered as the possible candidate of the double-charm molecular pentaquark with single strangeness.

In fact, the $\Omega_c^{(*)} D_s^{(*)}$ channels can be coupled with the $\Xi_c^{(\prime,*)} D^{(*)}$ systems. Thus, we need to consider the coupled channel effect between the $\Xi_c^{(\prime,*)} D^{(*)}$ and $\Omega_c^{(*)} D_s^{(*)}$ channels in the realistic calculations. An example is provided below to illustrate this point. Actually, the threshold of the $\Omega_c D_s$ channel is only 80 MeV higher than that of the $\Xi_c^{\prime} D^{*}$ channel, and we test the impact of the coupled channel effect between the $\Xi_c^{\prime} D^{*}$ and $\Omega_c D_s$ channels in the formation of the $\Xi_c^{\prime} D^{*}$ bound state with $I(J^P)=0(1/2^-)$. In Table~\ref{mixing}, we give the bound state properties for the $\Xi_c^{\prime} D^{*}$ state with $I(J^P)=0(1/2^-)$ without and with consideration of the coupled channel effect between the $\Xi_c^{\prime} D^{*}$ and $\Omega_c D_s$ channels. From the Table~\ref{mixing}, when taking into account the coupled channel effect between the $\Xi_c^{\prime} D^{*}$ and $\Omega_c D_s$ channels, the change of the binding energy for the $\Xi_c^{\prime} D^{*}$ state with $I(J^P)=0(1/2^-)$ is less than 0.8 MeV, and the contribution of the $\Omega_c D_s$ channel is less than 0.3 percent. Thus, the coupled channel effect between the $\Xi_c^{\prime} D^{*}$ and $\Omega_c D_s$ channels does not significantly affect the bound state properties for the $\Xi_c^{\prime} D^{*}$ state with $I(J^P)=0(1/2^-)$.
\renewcommand\tabcolsep{0.10cm}
\renewcommand{\arraystretch}{1.50}
\begin{table}[!htbp]
\caption{Bound state properties for the $\Xi_c^{\prime} D^{*}$ state with $I(J^P)=0(1/2^-)$ are analyzed in two cases: (I) the single channel analysis and (II) the coupled channel analysis between the $\Xi_c^{\prime} D^{*}$ and $\Omega_c D_s$ channels.}\label{mixing}
\centering
\begin{tabular}{c|cc|ccc}\toprule[1.0pt]\toprule[1.0pt]
Cases&\multicolumn{2}{c|}{I}&\multicolumn{3}{c}{II}\\\midrule[1.0pt]
$\Lambda(\rm{GeV})$ &$E(\rm {MeV})$ &$r_{\rm RMS}(\rm {fm})$ &$E(\rm {MeV})$ &$r_{\rm RMS}(\rm {fm})$ &$P(\Xi_c^{\prime} D^{*}/\Omega_c D_s)$ \\\midrule[1.0pt]
$0.92$&$-0.66$&$3.62$&$-0.90$ &3.20&\textbf{99.90}/0.10\\
$0.96$&$-4.94$&$1.52$&$-5.51$&1.45&\textbf{99.78}/0.22\\
$1.00$&$-13.74$&$0.99$&$-14.45$ &0.97&\textbf{99.74}/0.26\\
\bottomrule[1pt]\bottomrule[1pt]
\end{tabular}
\end{table}

In our previous discussions, we have systematically investigated the bound state properties of the $\Xi_c^{(\prime,*)} D^{(*)}$ systems by taking into account various effects, like the $S$-$D$ wave mixing effect and the coupled channel effect. Our findings indicate the existence of the ten most promising candidates of the double-charm molecular pentaquarks with single strangeness, including the $\Xi_c D$ state with $I(J^P)=0(1/2^-)$, the $\Xi_c^{\prime} D$ state with $I(J^P)=0(1/2^-)$, the $\Xi_c D^{*}$ states with $I(J^P)=0(1/2^-,3/2^-)$, the $\Xi_c^{*} D$ state with $I(J^P)=0(3/2^-)$, the $\Xi_c^{\prime} D^{*}$ states with $I(J^P)=0(1/2^-,3/2^-)$, and the $\Xi_c^{*} D^{*}$ states with $I(J^P)=0(1/2^-,3/2^-,5/2^-)$, which is in agreement with the theoretical prediction presented in Ref. \cite{Dong:2021bvy}. Furthermore, the $\Xi_c^{*} D^{*}$ state with $I(J^P)=1(5/2^-)$ can be regarded as the possible candidate of the double-charm molecular pentaquark with single strangeness.

\section{The M1 radiative decay behaviors and the magnetic moments}\label{sec3}

By analyzing the mass spectra of the $\Xi_c^{(\prime,*)} D^{(*)}$ systems, we have determined that the $\Xi_c D$ state with $I(J^P)=0(1/2^-)$, the $\Xi_c^{\prime} D$ state with $I(J^P)=0(1/2^-)$, the $\Xi_c D^{*}$ states with $I(J^P)=0(1/2^-,3/2^-)$, the $\Xi_c^{*} D$ state with $I(J^P)=0(3/2^-)$, the $\Xi_c^{\prime} D^{*}$ states with $I(J^P)=0(1/2^-,3/2^-)$, and the $\Xi_c^{*} D^{*}$ states with $I(J^P)=0(1/2^-,3/2^-,5/2^-)$ can be considered as the most promising candidates of the double-charm molecular pentaquarks with single strangeness, which can provide the important information for the experimental construction of the family of the $\Xi_c^{(\prime,*)} D^{(*)}$-type double-charm molecular pentaquarks with single strangeness. Nevertheless, the determination of their spin-parity quantum numbers is the important research topic if the relevant structures are observed within the corresponding energy regions in the future experiments. Furthermore, these molecular pentaquarks and the conventional $\Omega_{cc}$ baryons \cite{Yu:2022lel} may share the same quantum number and similar mass. For instance, the $\Xi_c^{\prime} D$ state with $I(J^P)=0(1/2^-)$ and the $\Omega_{cc}(1/2^-,2P)$ state, the $\Xi_c D^{*}$ state with $I(J^P)=0(3/2^-)$ and the $\Omega_{cc}(3/2^-,2P)$ state, the $\Xi_c^{\prime} D^{*}$ state with $I(J^P)=0(1/2^-)$ and the $\Omega_{cc}(1/2^-,3P)$ state, the $\Xi_c^{\prime} D^{*}$ state with $I(J^P)=0(3/2^-)$ and the  $\Omega_{cc}(3/2^-,3P)$ state, and so on \cite{Yu:2022lel}.  Thus, the construction of the family of the $\Xi_c^{(\prime,*)} D^{(*)}$-type double-charm molecular pentaquarks with single strangeness will face the significant challenges in the future experiments. To identify the spin-parity quantum numbers and the configurations of the $\Xi_c^{(\prime,*)} D^{(*)}$-type double-charm molecular pentaquarks with single strangeness experimentally, more physical observables need to be discussed based on the obtained mass spectra and spatial wave functions.

It is widely recognised that the electromagnetic properties of the hadron can indicate its inner structure, which is crucial for distinguishing the spin-parity quantum numbers and the configurations of the hadrons. This section concentrates on exploring the electromagnetic properties of our obtained isoscalar $\Xi_c^{(\prime,*)} D^{(*)}$-type double-charm molecular pentaquarks with single strangeness, and the constituent quark model is employed in our concrete calculations, which has been extensively employed in investigating the electromagnetic properties of the hadrons over the past few years \cite{Liu:2003ab,Huang:2004tn,Zhu:2004xa,Haghpayma:2006hu,Wang:2016dzu,Deng:2021gnb,Gao:2021hmv,Zhou:2022gra,Wang:2022tib,Li:2021ryu,Schlumpf:1992vq,Schlumpf:1993rm,Cheng:1997kr,Ha:1998gf,Ramalho:2009gk,Girdhar:2015gsa,Menapara:2022ksj,Mutuk:2021epz,Menapara:2021vug,Menapara:2021dzi,Gandhi:2018lez,Dahiya:2018ahb,Kaur:2016kan,Thakkar:2016sog,Shah:2016vmd,Dhir:2013nka,Sharma:2012jqz,Majethiya:2011ry,Sharma:2010vv,Dhir:2009ax,Simonis:2018rld,Ghalenovi:2014swa,Kumar:2005ei,Rahmani:2020pol,Hazra:2021lpa,Gandhi:2019bju,Majethiya:2009vx,Shah:2016nxi,Shah:2018bnr,Ghalenovi:2018fxh,Wang:2022nqs,Mohan:2022sxm,An:2022qpt,Kakadiya:2022pin,Wu:2022gie,Wang:2023bek,Wang:2023aob}.

\subsection{Formalism}

In this work, we investigate the electromagnetic properties including the M1 radiative decay widths and the magnetic moments of our obtained isoscalar $\Xi_c^{(\prime,*)} D^{(*)}$ molecular candidates.  In the following, we focus on how to calculate the M1 radiative decay widths and the magnetic moments of our obtained isoscalar $\Xi_c^{(\prime,*)} D^{(*)}$ molecular candidates based on the constituent quark model.

As shown in previous studies \cite{Dey:1994qi,Simonis:2018rld,Gandhi:2019bju,Hazra:2021lpa,Li:2021ryu,Zhou:2022gra,Wang:2022tib,Rahmani:2020pol,Menapara:2022ksj,Menapara:2021dzi,Gandhi:2018lez,Majethiya:2011ry,Majethiya:2009vx,Shah:2016nxi,Ghalenovi:2018fxh,Wang:2022nqs,Mohan:2022sxm,An:2022qpt,Kakadiya:2022pin,Wang:2023bek,Wang:2023aob}, the determination of the M1 radiative decay widths of the hadronic states can be achieved through analysis of the corresponding transition magnetic moments. For the $H \to H^{\prime} \gamma$ process, there exists the relation between the M1 radiative decay width, which is denoted as $\Gamma_{H \to H^{\prime}\gamma}$, and the corresponding transition magnetic moment, which is marked as $\mu_{H \to H^{\prime}}$ \cite{Wang:2022nqs,Wang:2023bek,Wang:2023aob}, i.e.,
\begin{eqnarray}
 \Gamma_{H \to H^{\prime}\gamma}= \frac{\alpha_{\rm {EM}}}{2J_{H}+1}\frac{k^{3}}{m_{p}^{2}}\frac{\sum\limits_{J_{H^{\prime}z},J_{Hz}}\left(\begin{array}{ccc} J_{H^{\prime}}&1&J_{H}\\-J_{H^{\prime}z}&0&J_{Hz}\end{array}\right)^2}{\left(\begin{array}{ccc} J_{H^{\prime}}&1&J_{H}\\-J_{z}&0&J_{z}\end{array}\right)^2}\frac{\left|\mu_{H \to H^{\prime}}\right|^2}{\mu_N^2}.\nonumber\\
\end{eqnarray}
Here, the emitted photon's momentum is represented by $k$, which can be calculated using the formula $k=(m_{H}^2-m_{H^{\prime}}^2)/2m_{H}$. The proton's mass is labeled as $m_p$ and has a value of $938~{\rm{MeV}}$ \cite{ParticleDataGroup:2022pth}, and the electromagnetic fine structure constant, denoted by $\alpha_{\rm {EM}}$, has an approximate value of ${1}/{137}$. The lowest value between $J_H$ and $J_{H^{\prime}}$ is designated as $J_z$, and we indicate the 3-$j$ coefficient through the notation $\left(\begin{array}{ccc} a&b&c\\d&e&f\end{array}\right)$.

The isoscalar $\Xi_c^{(\prime,*)} D^{(*)}$ molecular candidates have not yet been found experimentally \cite{Liu:2013waa,Hosaka:2016pey,Chen:2016qju,Richard:2016eis,Lebed:2016hpi,Brambilla:2019esw,Liu:2019zoy,Chen:2022asf,Olsen:2017bmm,Guo:2017jvc,Meng:2022ozq}. However, their M1 radiative decay behaviors are dependent on their binding energies \cite{Wang:2022nqs,Wang:2023bek,Wang:2023aob}. For simplicity, we adopt the same binding energies for the initial and final $\Xi_c^{(\prime,*)} D^{(*)}$ molecular states to analyze their M1 radiative decay behaviors in this work. Based on the aforementioned assumption, we can evaluate the transition magnetic moments between our obtained isoscalar $\Xi_c^{(\prime,*)} D^{(*)}$ molecular candidates by calculating the following relation \cite{Li:2021ryu,Zhou:2022gra,Wang:2022tib,Wang:2022nqs,Majethiya:2009vx,Majethiya:2011ry,Shah:2016nxi,Gandhi:2018lez,Simonis:2018rld,Ghalenovi:2018fxh,Gandhi:2019bju,Rahmani:2020pol,Hazra:2021lpa,Menapara:2021dzi,Menapara:2022ksj,Kakadiya:2022pin,Wang:2023bek,Wang:2023aob}
\begin{eqnarray}
\mu_{H \to H^{\prime}}=\left\langle{J_{H^{\prime}},J_{z}\left|\sum_{j}\hat{\mu}_{zj}^{\rm spin}e^{-i {\bf k}\cdot{\bf r}_j}+\hat{\mu}_z^{\rm orbital}\right|J_{H},J_{z}}\right\rangle.
\end{eqnarray}
In the above expression, the exponential factor $e^{-i {\bf k}\cdot{\bf r}_j}$ is the spatial wave function of the emitted photon,  ${\bf k}$ refers to the momentum of the emitted photon, and ${\bf r}_j$ stands for the coordinate of the $j$-th quark. Here, we provide the definition of the $z$-component for both the spin magnetic moment operator and the orbital magnetic moment operator as follows \cite{Liu:2003ab,Huang:2004tn,Zhu:2004xa,Haghpayma:2006hu,Wang:2016dzu,Deng:2021gnb,Gao:2021hmv,Zhou:2022gra,Wang:2022tib,Li:2021ryu,Schlumpf:1992vq,Schlumpf:1993rm,Cheng:1997kr,Ha:1998gf,Ramalho:2009gk,Girdhar:2015gsa,Menapara:2022ksj,Mutuk:2021epz,Menapara:2021vug,Menapara:2021dzi,Gandhi:2018lez,Dahiya:2018ahb,Kaur:2016kan,Thakkar:2016sog,Shah:2016vmd,Dhir:2013nka,Sharma:2012jqz,Majethiya:2011ry,Sharma:2010vv,Dhir:2009ax,Simonis:2018rld,Ghalenovi:2014swa,Kumar:2005ei,Rahmani:2020pol,Hazra:2021lpa,Gandhi:2019bju,Majethiya:2009vx,Shah:2016nxi,Shah:2018bnr,Ghalenovi:2018fxh,Wang:2022nqs,Mohan:2022sxm,An:2022qpt,Kakadiya:2022pin,Wu:2022gie,Wang:2023bek,Wang:2023aob}
\begin{eqnarray}
\hat{\mu}_{zj}^{\rm spin}&=&\frac{e_j}{2m_j}\hat{\sigma}_{zj},\\
\hat{\mu}_z^{\rm orbital}&=&\left(\frac{m_{m}}{m_{b}+m_{m}}\frac{e_b}{2m_b}+\frac{m_{b}}{m_{b}+m_{m}}\frac{e_m}{2m_m}\right)\hat{L}_z,
\end{eqnarray}
where the mass, the electric charge, and the $z$-component of the Pauli spin operator of the $j$-th constituent of the hadronic state are indicated as $m_j$, $e_j$, and $\hat{\sigma}_{zj}$, respectively. To differentiate between the baryons and the mesons within the hadronic molecular states, the subscripts $b$ and $m$ are used. Furthermore, the $z$-component of the orbital angular momentum operator between the meson and the baryon in the hadronic molecular state is denoted by $\hat{L}_z$.

In addition, we can calculate the magnetic moments, which are denoted as $\mu_{H}$, of our obtained isoscalar $\Xi_c^{(\prime,*)} D^{(*)}$ molecular candidates by taking the following equation \cite{Liu:2003ab,Huang:2004tn,Zhu:2004xa,Haghpayma:2006hu,Wang:2016dzu,Deng:2021gnb,Gao:2021hmv,Zhou:2022gra,Wang:2022tib,Li:2021ryu,Schlumpf:1992vq,Schlumpf:1993rm,Cheng:1997kr,Ha:1998gf,Ramalho:2009gk,Girdhar:2015gsa,Menapara:2022ksj,Mutuk:2021epz,Menapara:2021vug,Menapara:2021dzi,Gandhi:2018lez,Dahiya:2018ahb,Kaur:2016kan,Thakkar:2016sog,Shah:2016vmd,Dhir:2013nka,Sharma:2012jqz,Majethiya:2011ry,Sharma:2010vv,Dhir:2009ax,Simonis:2018rld,Ghalenovi:2014swa,Kumar:2005ei,Rahmani:2020pol,Hazra:2021lpa,Gandhi:2019bju,Majethiya:2009vx,Shah:2016nxi,Shah:2018bnr,Ghalenovi:2018fxh,Wang:2022nqs,Mohan:2022sxm,An:2022qpt,Kakadiya:2022pin,Wu:2022gie,Wang:2023bek,Wang:2023aob}
\begin{eqnarray}
\mu_{H}&=&\left\langle{J_{H},J_{H}\left|\sum_{j}\hat{\mu}_{zj}^{\rm spin}+\hat{\mu}_z^{\rm orbital}\right|J_{H},J_{H}}\right\rangle.
\end{eqnarray}

When studying the transition magnetic moments and the magnetic moments of the isoscalar $\Xi_c^{(\prime,*)} D^{(*)}$ molecular candidates based on the constituent quark model, it is crucial to account for the spatial wave functions of both the molecular states and the corresponding constituent hadrons as the inputs \cite{Wang:2022nqs,Wang:2023bek,Wang:2023aob}. For the isoscalar $\Xi_c^{(\prime,*)} D^{(*)}$ molecular candidates, we use the precise spatial wave functions obtained by numerically solving the coupled channel Schr\"odinger equation in the previous section. For both the baryons $\Xi_c^{(\prime,*)}$ and the mesons $D^{(*)}$, their spatial wave functions are described using the simple harmonic oscillator (SHO) wave function \cite{Wang:2022nqs,Wang:2023bek,Wang:2023aob}, i.e.,
\begin{eqnarray}
\phi_{n,l,m}(\beta,{\bf r})&=&\sqrt{\frac{2n!}{\Gamma(n+l+\frac{3}{2})}}L_{n}^{l+\frac{1}{2}}(\beta^2r^2)\beta^{l+\frac{3}{2}}{\mathrm e}^{-\frac{\beta^2r^2}{2}}r^l Y_{l m}(\Omega_{\bf r}),\nonumber\\
\end{eqnarray}
where the quantum numbers $n$, $l$, and $m$ are utilized to denote the radial, orbital, and magnetic characteristics of the hadron, respectively. To represent the associated Laguerre polynomial and the spherical harmonic function, we use the notations $L_{n}^{l+\frac{1}{2}}(x)$ and $Y_{l m}(\Omega_{\bf r})$, respectively. The parameters in the SHO wave functions are denoted by $\beta$, which can be estimated by fitting their mass spectra \cite{ParticleDataGroup:2022pth}. In the specific calculations, we choose $(\beta_{\rho},\beta_{\lambda})_{\Xi_c}=(0.301,0.383)$, $(\beta_{\rho},\beta_{\lambda})_{\Xi_c^{\prime}}=(0.252,0.383)$, $(\beta_{\rho},\beta_{\lambda})_{\Xi_c^{*}}=(0.243,0.358)$, $\beta_{D}=0.357$, and $\beta_{D^*}=0.307$ \cite{Luo:2023sne}, which are expressed in the units of GeV. Furthermore, it is necessary to use the spherical Bessel function $j_l(x)$ and the spherical harmonic function $Y_{l m}(\Omega)$ to expand the spatial wave function of the emitted photon $e^{-i{\bf k}\cdot{\bf r}_j}$ \cite{Khersonskii:1988krb}, i.e., $e^{-i{\bf k}\cdot{\bf r}_j}=\sum\limits_{l=0}^\infty\sum\limits_{m=-l}^l4\pi(-i)^lj_l(kr_j)Y_{lm}^*(\Omega_{\bf k})Y_{lm}(\Omega_{{\bf r}_j})$,
which can be used to calculate the contribution of the spatial wave functions of the initial and final states $\left\langle \phi_f \left|e^{-i {\bf k}\cdot{\bf r}_j}\right| \phi_i\right\rangle$.

\subsection{The transition magnetic moments and the magnetic moments of the baryons $\Xi_c^{(\prime,*)}$ and the mesons $D^{(*)}$}

Within the constituent quark model, the transition magnetic moment of the hadronic molecular state can be expressed as the linear combination of the transition magnetic moments and the magnetic moments of the corresponding constituent hadrons \cite{Wang:2022nqs,Wang:2023bek,Wang:2023aob}. Nevertheless, no experimental data are available on the transition magnetic moments and the magnetic moments of the baryons $\Xi_c^{(\prime,*)}$ and the mesons $D^{(*)}$ at present \cite{ParticleDataGroup:2022pth}. Thus, we first study the transition magnetic moments and the magnetic moments of the baryons $\Xi_c^{(\prime,*)}$ and the mesons $D^{(*)}$ based on the constituent quark model. In this model, the quark masses serve as the essential input parameters, and the values of $m_{u}=0.336\,\mathrm{GeV}$, $m_{d}=0.336\,\mathrm{GeV}$, $m_{s}=0.450\,\mathrm{GeV}$, and $m_{c}=1.680\,\mathrm{GeV}$ are utilized to conduct quantitative analyses of the hadronic transition magnetic moments and magnetic moments \cite{Kumar:2005ei,Li:2021ryu,Zhou:2022gra,Wang:2022tib,Wang:2022nqs,Wang:2023aob}.

The flavour wave functions for the baryons $\Xi_c^{(\prime,*)}$ and the mesons $D^{(*)}$ can be constructed in the following manner
\begin{eqnarray*}
\begin{array}{ll}
\Xi_c^{+}=\dfrac{1}{\sqrt{2}}\left(usc-suc\right),&~~~~~~~
\Xi_c^{0}=\dfrac{1}{\sqrt{2}}\left(dsc-sdc\right),\\
\Xi_c^{\prime(*)+}=\dfrac{1}{\sqrt{2}}\left(usc+suc\right),&~~~~~~~
\Xi_c^{\prime(*)0}=\dfrac{1}{\sqrt{2}}\left(dsc+sdc\right),\\
D^{(*)0}=c \bar u,&~~~~~~~
D^{(*)+}=c \bar d,
\end{array}
\end{eqnarray*}
and the spin wave functions $|S, m_{S}\rangle$ corresponding to them can be constructed as
\begin{eqnarray*}
\Xi_c&:&\left\{
  \begin{array}{l}
    \left|\dfrac{1}{2},\dfrac{1}{2}\right\rangle=\dfrac{1}{\sqrt{2}}\left(\uparrow\downarrow\uparrow-\downarrow\uparrow\uparrow\right)\\
    \left|\dfrac{1}{2},-\dfrac{1}{2}\right\rangle=\dfrac{1}{\sqrt{2}}\left(\uparrow\downarrow\downarrow-\downarrow\uparrow\downarrow\right)
  \end{array}
\right.,\\
\Xi_c^{\prime}&:&\left\{
  \begin{array}{l}
    \left|\dfrac{1}{2},\dfrac{1}{2}\right\rangle=\dfrac{1}{\sqrt{6}}\left(2\uparrow\uparrow\downarrow-\downarrow\uparrow\uparrow-\uparrow\downarrow\uparrow\right)\\
    \left|\dfrac{1}{2},-\dfrac{1}{2}\right\rangle=\dfrac{1}{\sqrt{6}}\left(\downarrow\uparrow\downarrow+\uparrow\downarrow\downarrow-2\downarrow\downarrow\uparrow\right)
  \end{array}
\right.,\\
\Xi_c^{*}&:&\left\{
  \begin{array}{l}
    \left|\dfrac{3}{2},\dfrac{3}{2}\right\rangle=\uparrow\uparrow\uparrow\\
    \left|\dfrac{3}{2},\dfrac{1}{2}\right\rangle=\dfrac{1}{\sqrt{3}}\left(\downarrow\uparrow\uparrow+\uparrow\downarrow\uparrow+\uparrow\uparrow\downarrow\right)\\
     \left|\dfrac{3}{2},-\dfrac{1}{2}\right\rangle=\dfrac{1}{\sqrt{3}}\left(\downarrow\downarrow\uparrow+\uparrow\downarrow\downarrow+\downarrow\uparrow\downarrow\right)\\
    \left|\dfrac{3}{2},-\dfrac{3}{2}\right\rangle=\downarrow\downarrow\downarrow
  \end{array}
\right.,\\
D&:&
    \left|0,0\right\rangle=\dfrac{1}{\sqrt{2}}\left(\uparrow\downarrow-\downarrow\uparrow\right),\\
 D^{*}&:&\left\{
  \begin{array}{l}
    \left|1,1\right\rangle=\uparrow\uparrow\\
    \left|1,0\right\rangle=\dfrac{1}{\sqrt{2}}\left(\uparrow\downarrow+\downarrow\uparrow\right)\\
    \left|1,-1\right\rangle=\downarrow\downarrow
  \end{array}
\right.,
\end{eqnarray*}
where the spins and the corresponding third components of these hadrons are denoted as the notations $S$ and $m_{S}$ in the spin wave functions $|S, m_{S}\rangle$, respectively. In addition, the third components of the spins of the quarks, which have values of $+{1}/{2}$ and $-{1}/{2}$, are represented by the notations $\uparrow$ and $\downarrow$, respectively.

Table~\ref{MTT1} presents our obtained results for the transition magnetic moments and the magnetic moments of the $\Xi_c^{(\prime,*)}$ baryons and the $D^{(*)}$ mesons, where we also compare our findings with those of the previous studies.

\renewcommand\tabcolsep{0.33cm}
\renewcommand{\arraystretch}{1.50}
\begin{table}[!htbp]
  \caption{Our obtained transition magnetic moments and magnetic moments of the baryons $\Xi_c^{(\prime,*)}$ and the mesons $D^{(*)}$, and comparison with the values obtained from other studies. Both the transition magnetic moments and the magnetic moments of the hadrons are expressed in units of $\mu_N=e/2m_p$.}\label{MTT1}
\begin{tabular}{c|c|l}
\toprule[1.0pt]
\toprule[1.0pt]
\multicolumn{3}{c}{Transition magnetic moments}\\\midrule[1.0pt]
Decay processes &  \multicolumn{1}{c|}{Our work}  &  \multicolumn{1}{c}{Other works} \\\hline
$\Xi^{\prime+}_c \to \Xi^{+}_c \gamma$ & $-1.412$ &$-1.428$ \cite{Hazra:2021lpa},\,$-1.390$ \cite{Kumar:2005ei}\\
$\Xi^{\prime0}_c \to \Xi^{0}_c \gamma$ & $0.127$ &$0.136$ \cite{Wang:2022tib},\,$0.130$ \cite{Kumar:2005ei}\\
$\Xi^{*+}_c \to \Xi^{+}_c \gamma$ & $1.904$ &$1.940$ \cite{Kumar:2005ei},\,$1.960$ \cite{Simonis:2018rld}\\
$\Xi^{*0}_c \to \Xi^{0}_c \gamma$ & $-0.163$ &$-0.176$ \cite{Gandhi:2019bju},\,$-0.160$ \cite{Aliev:2009jt}\\
$\Xi^{*+}_c \to \Xi^{\prime+}_c \gamma$ & $0.191$ &$0.199$ \cite{Wang:2022nqs},\,$0.170$ \cite{Kumar:2005ei}\\
$\Xi^{*0}_c \to \Xi^{\prime0}_c \gamma$ & $-1.106$ &$-1.070$ \cite{Simonis:2018rld},\,$-1.030$ \cite{Simonis:2018rld}\\
$D^{*0} \to D^{0} \gamma$ & $2.140$ &$2.134$ \cite{Wang:2023bek},\,$2.233$ \cite{Zhou:2022gra}\\
$D^{*+} \to D^{+} \gamma$ & $-0.525$ &$-0.559$ \cite{Zhou:2022gra},\,$-0.540$ \cite{Simonis:2016pnh}\\\midrule[1.0pt]
\multicolumn{3}{c}{Magnetic moments}\\\midrule[1.0pt]
Hadrons &  \multicolumn{1}{c|}{Our work}  &  \multicolumn{1}{c}{Other works} \\\hline
$\Xi^{+}_c$ & $0.372$ &$0.372$ \cite{Wang:2022tib},\,$0.370$ \cite{Kumar:2005ei}\\
$\Xi^{0}_c$ & $0.372$ &$0.372$ \cite{Wang:2022tib},\,$0.366$ \cite{Kumar:2005ei}\\
$\Xi^{\prime+}_c$ & $0.654$&$0.654$ \cite{Wang:2022nqs},\,$0.650$ \cite{Glozman:1995xy}\\
$\Xi^{\prime0}_c$ & $-1.208$ &$-1.208$ \cite{Wang:2022nqs},\,$-1.200$ \cite{Zhang:2021yul}\\
$\Xi^{*+}_c$ & $1.539$ &$1.539$ \cite{Wang:2022nqs} ,\,$1.510$ \cite{Patel:2007gx}\\
$\Xi^{*0}_c$ & $-1.254$ &$-1.254$ \cite{Wang:2022nqs},\,$-1.260$ \cite{Barik:1983ics}\\
$D^{*0}$ & $-1.489$ &$-1.485$ \cite{Wang:2023bek},\,$-1.489$ \cite{Zhou:2022gra}\\
$D^{*+}$ & $1.303$ &$1.308$ \cite{Wang:2023bek},\,$1.303$ \cite{Zhou:2022gra}\\
\bottomrule[1.0pt]
\bottomrule[1.0pt]
\end{tabular}
\end{table}

Based on the numerical results presented in Table~\ref{MTT1}, it is found that our obtained transition magnetic moments and magnetic moments for the baryons $\Xi_c^{(\prime,*)}$ and the mesons $D^{(*)}$ are close to those obtained from other works, which indicates that our adopted quark masses in the constituent quark model are dependable. Thus, our subsequent results can provide the important theoretical guidance for the future experimental investigations regarding the electromagnetic properties of the isoscalar $\Xi_c^{(\prime,*)} D^{(*)}$ molecular pentaquarks.

\subsection{The M1 radiative decay behaviors of the isoscalar $\Xi_c^{(\prime,*)} D^{(*)}$ molecular pentaquarks}

Following the previous discussions, we now turn our attention to the electromagnetic properties including the M1 radiative decay behaviors and the magnetic moments of our obtained isoscalar $\Xi_c^{(\prime,*)} D^{(*)}$ molecular candidates.

For the isoscalar $\Xi_c^{(\prime,*)} D^{(*)}$ molecular states, the relevant flavour and spin-orbital wave functions have been constructed in the above section. Similar to the investigation of the mass spectra of the $\Xi_c^{(\prime,*)} D^{(*)}$ molecular states, we also consider the influences of the $S$-$D$ wave mixing effect and the coupled channel effect in the study of their M1 radiative decay behaviors and magnetic moments. To derive the transition magnetic moments and the magnetic moments of the $D$-wave channels, their spin-orbital wave functions $|{ }^{2 S+1} L_{J}\rangle$ need to be expanded through combining the spin wave function $|S,m_{S}\rangle$ and the orbital wave function $Y_{L m_{L}}$, i.e.,
\begin{eqnarray}
\left|{ }^{2 S+1} L_{J}\right\rangle=\sum_{m_{S},m_{L}} C_{S m_{S},L m_{L}}^{J M} |S,m_{S}\rangle Y_{L m_{L}}.
\end{eqnarray}

The transition magnetic moments and the M1 radiative decay behaviors of the hadronic molecular states depend on the bound energies of the initial and final hadronic molecular states \cite{Wang:2022nqs,Wang:2023bek,Wang:2023aob}, but these isoscalar $\Xi_c^{(\prime,*)} D^{(*)}$ molecular candidates have yet to be discovered experimentally \cite{Liu:2013waa,Hosaka:2016pey,Chen:2016qju,Richard:2016eis,Lebed:2016hpi,Brambilla:2019esw,Liu:2019zoy,Chen:2022asf,Olsen:2017bmm,Guo:2017jvc,Meng:2022ozq}. In this study, the same binding energies were utilised for both the initial and final $\Xi_c^{(\prime,*)} D^{(*)}$ molecular states and three typical values $-0.5$, $-6.0$, and $-12.0$ MeV were taken to explore their transition magnetic moments and M1 radiative decay widths. In Tables~\ref{TMM1} and \ref{TMM2}, we summarise the transition magnetic moments and the M1 radiative decay widths of our obtained isoscalar $\Xi_c^{(\prime,*)} D^{(*)}$ molecular candidates\footnote{In Tables~\ref{TMM1} and \ref{TMM2}, we labeled the radiative decay width less than $0.0005~\rm keV$ as $\mathcal{O}(0)$.}.

\renewcommand\tabcolsep{0.97cm}
\renewcommand{\arraystretch}{1.50}
\begin{table*}[!htbp]
  \caption{The transition magnetic moments $\mu_{H \to H^{\prime}}$ and the M1 radiative decay widths $\Gamma_{H \to H^{\prime}\gamma}$ of our obtained isoscalar $\Xi_c^{(\prime,*)} D^{(*)}$ molecular candidates. The results of the single channel analysis, the $S$-$D$ wave mixing analysis, and the coupled channel analysis are represented by Cases I, II, and III in the second column, respectively.}\label{TMM1}
\begin{tabular}{c|c|c|c}
\toprule[1.0pt]
\toprule[1.0pt]
Radiative decay processes&Cases& $\mu_{H \to H^{\prime}}(\mu_N)$&$\Gamma_{H \to H^{\prime}\gamma}({\rm keV})$ \\\midrule[1.0pt]
\multirow{2}{*}{$\Xi_c^{\prime}D\left|\frac{1}{2}^-\right\rangle \to \Xi_c D\left|\frac{1}{2}^-\right\rangle \gamma$}&I&$-0.550,\,-0.629,\,-0.635$&$3.168,\,4.135,\,4.212$\\
&III&$-0.417,\,-0.202,\,-0.015$&$1.817,\,0.425,\,0.002$\\
\multirow{2}{*}{$\Xi_cD^{*}\left|\frac{3}{2}^-\right\rangle \to \Xi_c D\left|\frac{1}{2}^-\right\rangle \gamma$}&I&$0.467,\,0.624,\,0.639$&$2.434,\,4.349,\,4.562$\\
&III&$0.316,\,0.199,\,0.110$&$1.113,\,0.441,\,0.134$\\
\multirow{2}{*}{$\Xi_cD^{*}\left|\frac{1}{2}^-\right\rangle \to \Xi_c D\left|\frac{1}{2}^-\right\rangle \gamma$}&I&$0.330,\,0.441,\,0.452$&$2.434,\,4.349,\,4.562$\\
&III&$0.314,\,0.379,\,0.385$&$2.206,\,3.197,\,3.305$\\
\multirow{2}{*}{$\Xi_c^{*}D\left|\frac{3}{2}^-\right\rangle \to \Xi_c D\left|\frac{1}{2}^-\right\rangle \gamma$}&I&$0.620,\,0.824,\,0.843$&$8.272,\,14.592,\,15.275$\\
&III&$0.591,\,0.525,\,0.281$&$7.503,\,5.919,\,1.698$\\
\multirow{1}{*}{$\Xi_c^{\prime} D^{*}\left|\frac{1}{2}^-\right\rangle \to \Xi_c D\left|\frac{1}{2}^-\right\rangle \gamma$}&III&$0.024,\,0.101,\,0.153$&$0.069,\,1.227,\,2.828$\\
\multirow{1}{*}{$\Xi_c^{\prime} D^{*}\left|\frac{3}{2}^-\right\rangle \to \Xi_c D\left|\frac{1}{2}^-\right\rangle \gamma$}&III&$-0.008,\,-0.031,\,-0.040$&$0.004,\,0.058,\,0.094$\\
\multirow{1}{*}{$\Xi_c^{*} D^{*}\left|\frac{1}{2}^-\right\rangle \to \Xi_c D\left|\frac{1}{2}^-\right\rangle \gamma$}&III&$0.013,\,0.017,\,0.029$&$0.038,\,0.070,\,0.201$\\
\multirow{1}{*}{$\Xi_c^{*} D^{*}\left|\frac{3}{2}^-\right\rangle \to \Xi_c D\left|\frac{1}{2}^-\right\rangle \gamma$}&III&$-0.003,\,-0.012,\,-0.056$&$0.001,\,0.016,\,0.373$\\
\multirow{1}{*}{$\Xi_c D^{*}\left|\frac{3}{2}^-\right\rangle \to \Xi_c^{\prime} D\left|\frac{1}{2}^-\right\rangle \gamma$}&III&$-0.068,\,-0.104,\,-0.088$&$0.001,\,0.001,\,0.001$\\
\multirow{1}{*}{$\Xi_c D^{*}\left|\frac{1}{2}^-\right\rangle \to \Xi_c^{\prime} D\left|\frac{1}{2}^-\right\rangle \gamma$}&III&$0.057,\,0.179,\,0.230$&$0.001,\,0.008,\,0.014$\\
\multirow{2}{*}{$\Xi_c^{*} D\left|\frac{3}{2}^-\right\rangle \to \Xi_c^{\prime} D\left|\frac{1}{2}^-\right\rangle \gamma$}&I&$-0.429,\,-0.453,\,-0.455$&$0.227,\,0.253,\,0.255$\\
&III&$-0.368,\,-0.151,\,-0.017$&$0.167,\,0.028,\,\mathcal{O}(0)$\\
\multirow{2}{*}{$\Xi_c^{\prime} D^{*}\left|\frac{1}{2}^-\right\rangle \to \Xi_c^{\prime} D\left|\frac{1}{2}^-\right\rangle \gamma$}&I&$0.336,\,0.442,\,0.452$&$2.523,\,4.371,\,4.567$\\
&III&$0.336,\,0.365,\,0.319$&$2.522,\,2.976,\,2.274$\\
\multirow{2}{*}{$\Xi_c^{\prime} D^{*}\left|\frac{3}{2}^-\right\rangle \to \Xi_c^{\prime} D\left|\frac{1}{2}^-\right\rangle \gamma$}&I&$0.462,\,0.610,\,0.627$&$2.380,\,4.161,\,4.391$\\
&III&$0.333,\,0.252,\,0.196$&$1.239,\,0.707,\,0.428$\\
\multirow{1}{*}{$\Xi_c^{*} D^{*}\left|\frac{1}{2}^-\right\rangle \to \Xi_c^{\prime} D\left|\frac{1}{2}^-\right\rangle \gamma$}&III&$-0.089,\,-0.201,\,-0.230$&$0.557,\,2.839,\,3.722$\\
\multirow{1}{*}{$\Xi_c^{*} D^{*}\left|\frac{3}{2}^-\right\rangle \to \Xi_c^{\prime} D\left|\frac{1}{2}^-\right\rangle \gamma$}&III&$0.088,\,0.196,\,0.220$&$0.273,\,1.344,\,1.703$\\
\multirow{1}{*}{$\Xi_c^{*} D\left|\frac{3}{2}^-\right\rangle \to \Xi_c D^{*}\left|\frac{3}{2}^-\right\rangle \gamma$}&III&$0.115,\,0.573,\,0.787$&$0.003,\,0.066,\,0.123$\\
\multirow{2}{*}{$\Xi_c^{\prime} D^{*}\left|\frac{1}{2}^-\right\rangle \to \Xi_c D^{*}\left|\frac{3}{2}^-\right\rangle \gamma$}&I&$-0.523,\,-0.592,\,-0.597$&$2.858,\,3.670,\,3.730$\\
&III&$-0.479,\,-0.438,\,-0.389$&$2.401,\,2.006,\,1.585$\\
\multirow{2}{*}{$\Xi_c^{\prime} D^{*}\left|\frac{3}{2}^-\right\rangle \to \Xi_c D^{*}\left|\frac{3}{2}^-\right\rangle \gamma$}&I&$-0.548,\,-0.619,\,-0.625$&$1.744,\,2.225,\,2.269$\\
&III&$-0.645,\,-0.865,\,-0.904$&$2.421,\,4.350,\,4.753$\\
\multirow{2}{*}{$\Xi_c^{*} D^{*}\left|\frac{1}{2}^-\right\rangle \to \Xi_c D^{*}\left|\frac{3}{2}^-\right\rangle \gamma$}&I&$-0.149,\,-0.194,\,-0.199$&$0.959,\,1.628,\,1.698$\\
&III&$-0.089,\,-0.055,\,-0.043$&$0.344,\,0.131,\,0.078$\\
\multirow{2}{*}{$\Xi_c^{*} D^{*}\left|\frac{3}{2}^-\right\rangle \to \Xi_c D^{*}\left|\frac{3}{2}^-\right\rangle \gamma$}&I&$-0.400,\,-0.521,\,-0.533$&$6.886,\,11.718,\,12.226$\\
&III&$-0.491,\,-0.702,\,-0.735$&$10.376,\,21.265,\,23.296$\\
\multirow{2}{*}{$\Xi_c^{*} D^{*}\left|\frac{5}{2}^-\right\rangle \to \Xi_c D^{*}\left|\frac{3}{2}^-\right\rangle \gamma$}&I&$0.472,\,0.623,\,0.634$&$7.988,\,13.955,\,14.435$\\
&III&$0.395,\,0.398,\,0.360$&$5.604,\,5.677,\,4.659$\\
\bottomrule[1.0pt]
\bottomrule[1.0pt]
\end{tabular}
\end{table*}

\renewcommand\tabcolsep{0.98cm}
\renewcommand{\arraystretch}{1.50}
\begin{table*}[!htbp]
  \caption{The transition magnetic moments $\mu_{H \to H^{\prime}}$ and the M1 radiative decay widths $\Gamma_{H \to H^{\prime}\gamma}$ of our obtained isoscalar $\Xi_c^{(\prime,*)} D^{(*)}$ molecular candidates. The results of the single channel analysis, the $S$-$D$ wave mixing analysis, and the coupled channel analysis are represented by Cases I, II, and III in the second column, respectively.}\label{TMM2}
\begin{tabular}{c|c|c|c}
\toprule[1.0pt]
\toprule[1.0pt]
Radiative decay processes&Cases& $\mu_{H \to H^{\prime}}(\mu_N)$&$\Gamma_{H \to H^{\prime}\gamma}({\rm keV})$ \\\midrule[1.0pt]
\multirow{1}{*}{$\Xi_c^{*}D\left|\frac{3}{2}^-\right\rangle \to \Xi_c D^{*}\left|\frac{1}{2}^-\right\rangle \gamma$}&III&$0.037,\,0.060,\,0.028$&$\mathcal{O}(0),\,0.001,\,\mathcal{O}(0)$\\
\multirow{2}{*}{$\Xi_c^{\prime}D^{*}\left|\frac{1}{2}^-\right\rangle \to \Xi_c D^{*}\left|\frac{1}{2}^-\right\rangle \gamma$}&I&$0.185,\,0.209,\,0.211$&$0.357,\,0.459,\,0.466$\\
&III&$0.148,\,0.084,\,0.042$&$0.228,\,0.075,\,0.019$\\
\multirow{2}{*}{$\Xi_c^{\prime}D^{*}\left|\frac{3}{2}^-\right\rangle \to \Xi_c D^{*}\left|\frac{1}{2}^-\right\rangle \gamma$}&I&$-0.516,\,-0.583,\,-0.589$&$1.395,\,1.780,\,1.815$\\
&III&$-0.423,\,-0.273,\,-0.206$&$0.937,\,0.391,\,0.222$\\
\multirow{2}{*}{$\Xi_c^{*}D^{*}\left|\frac{1}{2}^-\right\rangle \to \Xi_c D^{*}\left|\frac{1}{2}^-\right\rangle \gamma$}&I&$-0.422,\,-0.550,\,-0.561$&$7.674,\,13.025,\,13.584$\\
&III&$-0.392,\,-0.434,\,-0.403$&$6.620,\,8.126,\,7.002$\\
\multirow{2}{*}{$\Xi_c^{*}D^{*}\left|\frac{3}{2}^-\right\rangle \to \Xi_c D^{*}\left|\frac{1}{2}^-\right\rangle \gamma$}&I&$0.471,\,0.614,\,0.628$&$4.782,\,8.137,\,8.490$\\
&III&$0.428,\,0.457,\,0.414$&$3.956,\,4.500,\,3.684$\\
\multirow{1}{*}{$\Xi_c^{\prime}D^{*}\left|\frac{1}{2}^-\right\rangle \to \Xi_c^{*} D\left|\frac{3}{2}^-\right\rangle \gamma$}&III&$-0.044,\,-0.214,\,-0.280$&$0.006,\,0.151,\,0.259$\\
\multirow{1}{*}{$\Xi_c^{\prime}D^{*}\left|\frac{3}{2}^-\right\rangle \to \Xi_c^{*} D\left|\frac{3}{2}^-\right\rangle \gamma$}&III&$0.110,\,0.335,\,0.349$&$0.022,\,0.206,\,0.223$\\
\multirow{2}{*}{$\Xi_c^{*}D^{*}\left|\frac{1}{2}^-\right\rangle \to \Xi_c^{*} D\left|\frac{3}{2}^-\right\rangle \gamma$}&I&$-0.337,\,-0.443,\,-0.452$&$2.539,\,4.381,\,4.573$\\
&III&$-0.325,\,-0.277,\,-0.132$&$2.359,\,1.717,\,0.388$\\
\multirow{2}{*}{$\Xi_c^{*}D^{*}\left|\frac{3}{2}^-\right\rangle \to \Xi_c^{*} D\left|\frac{3}{2}^-\right\rangle \gamma$}&I&$0.451,\,0.594,\,0.607$&$2.531,\,4.378,\,4.572$\\
&III&$0.455,\,0.556,\,0.467$&$2.571,\,3.839,\,2.712$\\
\multirow{2}{*}{$\Xi_c^{*}D^{*}\left|\frac{5}{2}^-\right\rangle \to \Xi_c^{*} D\left|\frac{3}{2}^-\right\rangle \gamma$}&I&$0.354,\,0.472,\,0.480$&$2.339,\,4.145,\,4.290$\\
&III&$0.356,\,0.367,\,0.240$&$2.365,\,2.513,\,1.076$\\
\multirow{3}{*}{$\Xi_c^{*}D^{*}\left|\frac{1}{2}^-\right\rangle \to \Xi_c^{\prime} D^{*}\left|\frac{1}{2}^-\right\rangle \gamma$}&I&$0.288,\,0.303,\,0.303$&$0.204,\,0.225,\,0.227$\\
&II&$0.287,\,0.302,\,0.303$&$0.203,\,0.225,\,0.226$\\
&III&$0.293,\,0.316,\,0.319$&$0.211,\,0.245,\,0.250$\\
\multirow{3}{*}{$\Xi_c^{*}D^{*}\left|\frac{3}{2}^-\right\rangle \to \Xi_c^{\prime} D^{*}\left|\frac{1}{2}^-\right\rangle \gamma$}&I&$-0.322,\,-0.338,\,-0.339$&$0.128,\,0.141,\,0.142$\\
&II&$-0.320,\,-0.336,\,-0.338$&$0.126,\,0.139,\,0.140$\\
&III&$-0.313,\,-0.302,\,-0.284$&$0.120,\,0.113,\,0.100$\\
\multirow{3}{*}{$\Xi_c^{*}D^{*}\left|\frac{1}{2}^-\right\rangle \to \Xi_c^{\prime} D^{*}\left|\frac{3}{2}^-\right\rangle \gamma$}&I&$0.100,\,0.104,\,0.104$&$0.025,\,0.026,\,0.027$\\
&II&$0.100,\,0.105,\,0.104$&$0.025,\,0.027,\,0.027$\\
&III&$0.119,\,0.154,\,0.163$&$0.035,\,0.059,\,0.065$\\
\multirow{3}{*}{$\Xi_c^{*}D^{*}\left|\frac{3}{2}^-\right\rangle \to \Xi_c^{\prime} D^{*}\left|\frac{3}{2}^-\right\rangle \gamma$}&I&$0.270,\,0.279,\,0.279$&$0.100,\,0.106,\,0.106$\\
&II&$0.270,\,0.281,\,0.279$&$0.100,\,0.108,\,0.107$\\
&III&$0.268,\,0.258,\,0.244$&$0.098,\,0.091,\,0.082$\\
\multirow{3}{*}{$\Xi_c^{*}D^{*}\left|\frac{5}{2}^-\right\rangle \to \Xi_c^{\prime} D^{*}\left|\frac{3}{2}^-\right\rangle \gamma$}&I&$-0.331,\,-0.350,\,-0.352$&$0.225,\,0.251,\,0.254$\\
&II&$-0.330,\,-0.349,\,-0.351$&$0.223,\,0.250,\,0.252$\\
&III&$-0.295,\,-0.209,\,-0.167$&$0.179,\,0.090,\,0.057$\\
\bottomrule[1.0pt]
\bottomrule[1.0pt]
\end{tabular}
\end{table*}

For the isoscalar $\Xi_c^{(\prime,*)} D^{(*)}$ molecular candidates, their M1 radiative decay behaviors are dependent on their transition magnetic moments and phase spaces \cite{Zhou:2022gra,Wang:2022tib,Wang:2022nqs,Wang:2023bek,Wang:2023aob}, and this fact can be illustrated through the following two examples. First, the $\Xi_c^{*} D\left|{3}/{2}^-\right\rangle \to \Xi_c^{\prime} D\left|{1}/{2}^-\right\rangle \gamma$ and the $\Xi_c^{\prime} D^{*}\left|{3}/{2}^-\right\rangle \to \Xi_c^{\prime} D\left|{1}/{2}^-\right\rangle \gamma$ display similar transition magnetic moments, but the radiative decay width of the $\Xi_c^{*} D\left|{3}/{2}^-\right\rangle \to \Xi_c^{\prime} D\left|{1}/{2}^-\right\rangle \gamma$ is smaller than that of the $\Xi_c^{\prime} D^{*}\left|{3}/{2}^-\right\rangle \to \Xi_c^{\prime} D\left|{1}/{2}^-\right\rangle \gamma$, which is attributed to the fact that the phase space of the $\Xi_c^{*} D\left|{3}/{2}^-\right\rangle \to \Xi_c^{\prime} D\left|{1}/{2}^-\right\rangle \gamma$ is much smaller than that of the $\Xi_c^{\prime} D^{*}\left|{3}/{2}^-\right\rangle \to \Xi_c^{\prime} D\left|{1}/{2}^-\right\rangle \gamma$. Second, the $\Xi_c^{*} D^{*}\left|{1}/{2}^-\right\rangle \to \Xi_c D^{*}\left|{3}/{2}^-\right\rangle \gamma$ and the $\Xi_c^{*} D^{*}\left|{3}/{2}^-\right\rangle \to \Xi_c D^{*}\left|{3}/{2}^-\right\rangle \gamma$ have the same phase space. However, the M1 radiative decay width of the $\Xi_c^{*} D^{*}\left|{3}/{2}^-\right\rangle \to \Xi_c D^{*}\left|{3}/{2}^-\right\rangle \gamma$ is significantly larger than that of the $\Xi_c^{*} D^{*}\left|{1}/{2}^-\right\rangle \to \Xi_c D^{*}\left|{3}/{2}^-\right\rangle \gamma$, which is due to the $\Xi_c^{*} D^{*}\left|{3}/{2}^-\right\rangle \to \Xi_c D^{*}\left|{3}/{2}^-\right\rangle \gamma$ having a stronger transition magnetic moment.

For the isoscalar $\Xi_c^{(\prime,*)} D^{(*)}$ molecular candidates, their M1 radiative decay behaviors are dependent on the binding energies of the initial and final $\Xi_c^{(\prime,*)} D^{(*)}$ molecular states \cite{Wang:2022nqs,Wang:2023bek,Wang:2023aob}, which is because the fact that the transition magnetic moments of the isoscalar $\Xi_c^{(\prime,*)} D^{(*)}$ molecular candidates are influenced by the spatial wave functions of the initial and final $\Xi_c^{(\prime,*)} D^{(*)}$ molecular states. In particular, the M1 radiative decay behaviors of our obtained isoscalar $\Xi_c^{(\prime,*)} D^{(*)}$ molecular candidates are particularly affected by the changes in the binding energies of the corresponding initial and final $\Xi_c^{(\prime,*)} D^{(*)}$ molecular states during the coupled channel analysis, this is mainly because there exists the significant change for the spatial wave functions of the different coupled channels as the binding energies change, which can be found in the previous section in the study of the mass spectra of the isoscalar $\Xi_c^{(\prime,*)} D^{(*)}$ molecular candidates. Therefore, the future experimental measurements of the binding energies of the isoscalar $\Xi_c^{(\prime,*)} D^{(*)}$ molecular candidates can advance our comprehension of their M1 radiative decay behaviors.

In most cases, the $S$-$D$ wave mixing effect has the negligible effect on the M1 radiative decay widths of our obtained isoscalar $\Xi_c^{(\prime,*)} D^{(*)}$ molecular candidates, and the coupled channel effect plays the crucial role on the M1 radiative decay widths of our obtained isoscalar $\Xi_c^{(\prime,*)} D^{(*)}$ molecular candidates, which exhibit the behavior similar to that of their mass spectra. As the binding energies of the initial and final $\Xi_c^{(\prime,*)} D^{(*)}$ molecular states increase, the M1 radiative decay widths of the corresponding processes obtained through the coupled channel analysis diverge from those of the single channel analysis. This phenomenon can be attributed to the increasing contribution of other coupled channels as the binding energies of the initial and final $\Xi_c^{(\prime,*)} D^{(*)}$ molecular states increase.

The study of the M1 radiative decay behaviors of the isoscalar $\Xi_c^{(\prime,*)} D^{(*)}$ molecular candidates can provide the valuable information to reflect their inner structures and properties, which may offer the vital insights for the future experimental determination of their spin-parity quantum numbers. A notable dissimilarity is observed in the M1 radiative decay behaviors of the $\Xi_c^{*} D^{*}\left|{1}/{2}^-\right\rangle \to \Xi_c D^{*}\left|{3}/{2}^-\right\rangle \gamma$ and the $\Xi_c^{*} D^{*}\left|{3}/{2}^-\right\rangle \to \Xi_c D^{*}\left|{3}/{2}^-\right\rangle \gamma$, so this allows for the future experimental differentiation of the $\Xi_c^{*} D^{*}$ molecular state with $I(J^P)=0(1/2^-)$ and the $\Xi_c^{*} D^{*}$ molecular state with $I(J^P)=0(3/2^-)$ through the analysis of their M1 radiative decay behaviors. Many similar examples exist in the present work, like the $\Xi_c^{\prime}D^{*}\left|{1}/{2}^-\right\rangle \to \Xi_c D^{*}\left|{1}/{2}^-\right\rangle \gamma$ and the $\Xi_c^{\prime}D^{*}\left|{3}/{2}^-\right\rangle \to \Xi_c D^{*}\left|{1}/{2}^-\right\rangle \gamma$, the $\Xi_c^{*}D^{*}\left|{1}/{2}^-\right\rangle \to \Xi_c^{\prime} D^{*}\left|{1}/{2}^-\right\rangle \gamma$ and the $\Xi_c^{*}D^{*}\left|{1}/{2}^-\right\rangle \to \Xi_c^{\prime} D^{*}\left|{3}/{2}^-\right\rangle \gamma$, and so on.

Similarly to the M1 radiative decay behaviors of the $\Xi_c \bar D^{*}\left|{3}/{2}^-\right\rangle \to \Xi_c \bar D\left|{1}/{2}^- \right\rangle \gamma$ and $\Xi_c  \bar D^{*}\left|{1}/{2}^-\right\rangle \to \Xi_c \bar D\left|{1}/{2}^-\right\rangle \gamma$ processes \cite{Wang:2022tib}, the same M1 radiative decay behaviors are found in the $\Xi_cD^{*}\left|{3}/{2}^-\right\rangle \to \Xi_c D\left|{1}/{2}^- \right\rangle \gamma$ and $\Xi_cD^{*}\left|{1}/{2}^-\right\rangle \to \Xi_c D\left|{1}/{2}^-\right\rangle \gamma$ processes when limited to the single channel analysis,  which can be attributed to the same binding properties for the $\Xi_c D^{*}$ molecular state with $I(J^P)=0(1/2^-)$ and the $\Xi_c D^{*}$ molecular state with $I(J^P)=0(3/2^-)$ in the single channel analysis. Thus, it is challenging to distinguish the $\Xi_c D^{*}$ molecular state with $I(J^P)=0(1/2^-)$ and the $\Xi_c D^{*}$ molecular state with $I(J^P)=0(3/2^-)$ based on the study of the M1 radiative decay behaviors for the $\Xi_cD^{*}\left|{3}/{2}^-\right\rangle \to \Xi_c D\left|{1}/{2}^-\right\rangle \gamma$ and $\Xi_cD^{*}\left|{1}/{2}^-\right\rangle \to \Xi_c D\left|{1}/{2}^-\right\rangle \gamma$ processes in the single channel analysis. However, the M1 radiative decay behaviors of the $\Xi_cD^{*}\left|{3}/{2}^-\right\rangle \to \Xi_c D\left|{1}/{2}^-\right\rangle \gamma$ and $\Xi_cD^{*}\left|{1}/{2}^-\right\rangle \to \Xi_c D\left|{1}/{2}^-\right\rangle \gamma$ processes exhibit the difference by introducing the coupled channel effect. Consequently, this finding indicates that the $\Xi_c D^{*}$ molecular state with $I(J^P)=0(1/2^-)$ and the $\Xi_c D^{*}$ molecular state with $I(J^P)=0(3/2^-)$ can be distinguished based on the study of the M1 radiative decay behaviors for the $\Xi_cD^{*}\left|{3}/{2}^-\right\rangle \to \Xi_c D\left|{1}/{2}^-\right\rangle \gamma$ and $\Xi_cD^{*}\left|{1}/{2}^-\right\rangle \to \Xi_c D\left|{1}/{2}^-\right\rangle \gamma$ processes after accounting for the influence of the coupled channel effect.

When considering the same binding energies for the initial and final $\Xi_c^{(\prime,*)} D^{(*)}$ molecular states, the radiative decay widths involving the same system with different spin-parity quantum numbers for the initial and final states are zero, and this occurs as a result of the phase spaces for such processes being zero \cite{Zhou:2022gra,Wang:2022tib,Wang:2023bek}. In fact, there may be differences for the binding energies of the initial and final $\Xi_c^{(\prime,*)} D^{(*)}$ molecular states for such M1 radiative decay processes. However, their M1 radiative decay widths are strongly suppressed (refer to Table \ref{TMM3} for more information), which is due to the tiny phase spaces associated with such M1 radiative decay processes.

\renewcommand\tabcolsep{0.05cm}
\renewcommand{\arraystretch}{1.50}
\begin{table}[!htbp]
  \caption{The transition magnetic moments and the M1 radiative decay widths of our obtained isoscalar $\Xi_c^{(\prime,*)} D^{(*)}$ molecular candidates, which involve the same system with different spin-parity quantum numbers for the initial and final states. The transition magnetic moments are expressed in $\mu_N$ units, and the radiative decay widths are measured in ${\rm keV}$. The results of the single channel analysis, the $S$-$D$ wave mixing analysis, and the coupled channel analysis are represented by Cases I, II, and III in the second column, respectively. $\Gamma_{H \to H^{\prime}\gamma}^{\rm Max}$ indicates the maximum value of the M1 radiative decay width when varying the binding energies of the corresponding initial and final molecular states.}\label{TMM3}
\begin{tabular}{c|c|c|c}
\toprule[1.0pt]
\toprule[1.0pt]
Decay processes&Cases& $\mu_{H \to H^{\prime}}$&$\Gamma_{H \to H^{\prime}\gamma}^{\rm Max}$ \\\midrule[1.0pt]
\multirow{2}{*}{$\Xi_c D^{*}\left|\frac{1}{2}^-\right\rangle \to \Xi_c D^{*}\left|\frac{3}{2}^-\right\rangle \gamma$}&I&$0.395,\,0.395,\,0.395$&\multirow{2}{*}{$0.002$}\\
&III&$0.306,\,0.139,\,0.077$&\\
\multirow{3}{*}{$\Xi_c^{\prime}D^{*}\left|\frac{3}{2}^-\right\rangle \to \Xi_c^{\prime} D^{*}\left|\frac{1}{2}^-\right\rangle \gamma$}&I&$-0.215,\,-0.211,\,-0.210$&\multirow{3}{*}{$0.0006$}\\
&II&$-0.214,\,-0.211,\,-0.209$&\\
&III&$-0.252,\,-0.296,\,-0.296$&\\
\multirow{2}{*}{$\Xi_c^{*}D^{*}\left|\frac{3}{2}^-\right\rangle \to \Xi_c^{*} D^{*}\left|\frac{1}{2}^-\right\rangle \gamma$}&I&$0.140,\,0.140,\,0.140$&\multirow{2}{*}{$0.0001$}\\
&II&$0.139,\,0.139,\,0.139$&\\
\multirow{2}{*}{$\Xi_c^{*}D^{*}\left|\frac{5}{2}^-\right\rangle \to \Xi_c^{*} D^{*}\left|\frac{3}{2}^-\right\rangle \gamma$}&I&$0.091,\,0.089,\,0.088$&\multirow{2}{*}{$0.00009$}\\
&II&$0.090,\,0.088,\,0.087$&\\
\bottomrule[1.0pt]
\bottomrule[1.0pt]
\end{tabular}
\end{table}

\subsection{The magnetic moments of the isoscalar $\Xi_c^{(\prime,*)} D^{(*)}$ molecular pentaquarks}

In Table~\ref{MagneticmomentsM}, we list the magnetic moments of our obtained isoscalar $\Xi_c^{(\prime,*)} D^{(*)}$ molecular candidates. In the specific calculations, the relevant numerical results were obtained through the single channel analysis, the $S$-$D$ wave mixing analysis, and the coupled channel analysis, respectively. For the $S$-$D$ wave mixing analysis and the coupled channel analysis, we take the binding energies $-0.5$, $-6.0$, and $-12.0$ MeV for the isoscalar $\Xi_c^{(\prime,*)} D^{(*)}$ molecular candidates to discuss their magnetic moments.

\renewcommand\tabcolsep{0.08cm}
\renewcommand{\arraystretch}{1.50}
\begin{table}[!htbp]
  \caption{The magnetic moments of our obtained isoscalar $\Xi_c^{(\prime,*)} D^{(*)}$ molecular candidates, which are expressed in units of $\mu_N$. The results of the single channel analysis, the $S$-$D$ wave mixing analysis, and the coupled channel analysis are represented by Cases I, II, and III, respectively.}
  \label{MagneticmomentsM}
\begin{tabular}{c|ccc}
\toprule[1.0pt]
\toprule[1.0pt]
Hadrons &  Case I& Case II & Case III\\\midrule[1.0pt]
$\Xi_c D|\frac{1}{2}^-\rangle$ & $0.372$ & & $0.371,\,0.368,\,0.370$\\
$\Xi_c^{\prime} D|\frac{1}{2}^-\rangle$ & $-0.277$ & & $-0.151,\,0.061,\,0.147$\\
$\Xi_c D^*|\frac{3}{2}^-\rangle$ & $0.279$ & & $0.403,\,0.638,\,0.745$\\
$\Xi_c D^*|\frac{1}{2}^-\rangle$ & $-0.186$ & & $-0.111,\,0.043,\,0.100$\\
$\Xi_c^* D|\frac{3}{2}^-\rangle$ & $0.143$ & & $0.213,\,0.527,\,0.560$\\
$\Xi_c^{\prime} D^*|\frac{1}{2}^-\rangle$ & $0.030$ & $0.031,\,0.032,\,0.032$& $0.065,\,0.148,\,0.198$\\
$\Xi_c^{\prime} D^*|\frac{3}{2}^-\rangle$ & $-0.370$ & $-0.364,\,-0.363,\,-0.365$& $-0.245,\,0.026,\,0.102$\\
$\Xi_c^{*} D^*|\frac{1}{2}^-\rangle$ & $0.110$ & $0.111,\,0.111,\,0.111$& \\
$\Xi_c^{*} D^*|\frac{3}{2}^-\rangle$ & $0.067$ & $0.070,\,0.071,\,0.070$& \\
$\Xi_c^{*} D^*|\frac{5}{2}^-\rangle$ & $0.050$ & $0.052,\,0.053,\,0.052$& \\
\bottomrule[1.0pt]
\bottomrule[1.0pt]
\end{tabular}
\end{table}

Based on the numerical results listed in Table~\ref{MagneticmomentsM}, we find that
\begin{itemize}
  \item In most cases, the $S$-$D$ wave mixing effect has the negligible impact on the magnetic moments of our obtained isoscalar $\Xi_c^{(\prime,*)} D^{(*)}$ molecular candidates, and the coupled channel effect plays the crucial role on the magnetic moments of our obtained isoscalar $\Xi_c^{(\prime,*)} D^{(*)}$ molecular candidates.
  \item In the coupled channel analysis, it is evident that the magnetic moments of our obtained isoscalar $\Xi_c^{(\prime,*)} D^{(*)}$ molecular candidates depend on the corresponding binding energies. Therefore, we strongly recommend the experimental determination of the binding energies of our obtained isoscalar $\Xi_c^{(\prime,*)} D^{(*)}$ molecular candidates, which is crucial for enhancing the understanding of the magnetic moments of our obtained isoscalar $\Xi_c^{(\prime,*)} D^{(*)}$ molecular candidates.
  \item The magnetic moments of our obtained isoscalar $\Xi_c^{(\prime,*)} D^{(*)}$ molecular candidates can reflect their inner structures and properties, and the determination of the spin-parity quantum numbers of these molecular pentaquarks can be achieved experimentally through the study of their magnetic moments in the future, such as the $\Xi_c D^{*}$ state with $I(J^P)=0(1/2^-)$ and the $\Xi_c D^{*}$ state with $I(J^P)=0(3/2^-)$, the $\Xi_c^{\prime} D^{*}$ state with $I(J^P)=0(1/2^-)$ and the $\Xi_c^{\prime} D^{*}$ state with $I(J^P)=0(3/2^-)$, and so on.
\end{itemize}

\section{Summary}\label{sec4}

In the field of hadron physics, the search for the exotic hadronic matter presents a compelling and vital area of research. Through the accumulation of the experimental data, the candidates of the hidden-charm pentaquarks with strangeness $P_{\psi s}^{\Lambda}(4459)/P_{\psi s}^{\Lambda}(4338)$ and the double-charm tetraquark state $T_{cc}(3875)^+$ were announced by the LHCb Collaboration, which has led the researchers to believe that the family of the double-charm molecular pentaquarks may exist within the hadron spectroscopy. Within the OBE model through the incorporation of both the $S$-$D$ wave mixing effect and the coupled channel effect, the double-charm molecular pentaquark candidates including the $\Sigma_c^{(*)} D^{(*)}$, $\Xi_c^{(\prime,*)} D_s^{(*)}$, and $\Omega_c^{(*)} D_s^{(*)}$ types have been predicted \cite{Chen:2021kad,Yalikun:2023waw,Wang:2023aob}. In the present work, the $\Xi_c^{(\prime,*)} D^{(*)}$ molecular systems are the main research objects.

\begin{figure}[htbp]
  \includegraphics[width=0.48\textwidth]{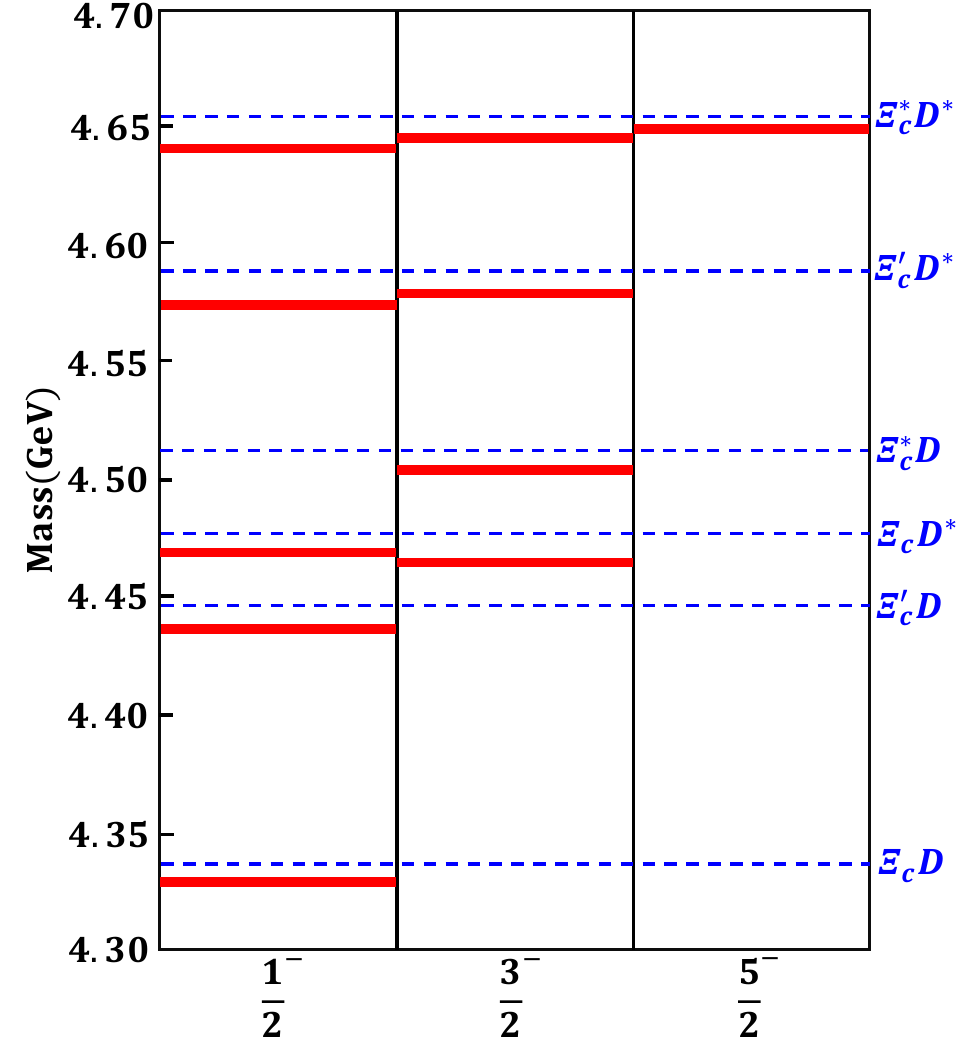}
  \caption{Our established mass spectra of the isoscalar $\Xi_c^{(\prime,*)} D^{(*)}$-type double-charm molecular pentaquark candidates with single strangeness. Here, the thresholds of the corresponding $\Xi_c^{(\prime,*)} D^{(*)}$ channels are plotted with the blue dotted lines, and the most promising candidates of the isoscalar $\Xi_c^{(\prime,*)} D^{(*)}$-type double-charm molecular pentaquarks with single strangeness are depicted with the red thick solid lines.}\label{Massspectra}
\end{figure}

First, we obtain the mass spectra and the corresponding spatial wave functions of the $\Xi_c^{(\prime,*)} D^{(*)}$-type double-charm molecular pentaquark candidates with single strangeness. Our calculations investigate the interactions between the $\Xi_c^{(\prime,*)} D^{(*)}$ systems by taking the OBE model and considering both the $S$-$D$ wave mixing effect and the coupled channel effect, and their bound state properties can then be analyzed by solving the coupled channel Schr$\ddot{\rm o}$dinger equation. As shown in Fig. \ref{Massspectra}, our investigation has revealed the existence of the ten most promising candidates of the double-charm molecular pentaquarks with single strangeness, including the $\Xi_c D$ state with $I(J^P)=0(1/2^-)$, the $\Xi_c^{\prime} D$ state with $I(J^P)=0(1/2^-)$, the $\Xi_c D^{*}$ states with $I(J^P)=0(1/2^-,3/2^-)$, the $\Xi_c^{*} D$ state with $I(J^P)=0(3/2^-)$, the $\Xi_c^{\prime} D^{*}$ states with $I(J^P)=0(1/2^-,3/2^-)$, and the $\Xi_c^{*} D^{*}$ states with $I(J^P)=0(1/2^-,3/2^-,5/2^-)$, which is in accordance with the theoretical analysis presented in Ref. \cite{Dong:2021bvy}. Moreover, the $\Xi_c D^{*}$ molecules with $I(J^P)=0(1/2^-,3/2^-)$ and the $\Xi_c \bar D^{*}$ molecules with $I(J^P)=0(1/2^-,3/2^-)$ \cite{Wang:2022mxy} display the similar spectral behavior, and the $\Xi_c D^{*}$ molecular states with $I(J^P)=0(1/2^-,3/2^-)$ can be separated into two states owing to the contribution of the coupled channel effect.

After that, we evaluate the M1 radiative decay behaviors and the magnetic moments of the isoscalar $\Xi_c^{(\prime,*)} D^{(*)}$ molecular states based on the obtained mass spectra and spatial wave functions. Our calculations employ the constituent quark model and account for both the $S$-$D$ wave mixing effect and the coupled channel effect. Based on our numerical calculations, we find that the study of the M1 radiative decay behaviors of the isoscalar $\Xi_c^{(\prime,*)} D^{(*)}$ molecules can yield the significant insights for the future experiments to determine their spin-parity quantum numbers, and the M1 radiative decay behaviors of the $\Xi_cD^{*}\left|{3}/{2}^-\right\rangle \to \Xi_c D\left|{1}/{2}^- \right\rangle \gamma$ and $\Xi_cD^{*}\left|{1}/{2}^-\right\rangle \to \Xi_c D\left|{1}/{2}^-\right\rangle \gamma$ processes are similar to those of the $\Xi_c \bar D^{*}\left|{3}/{2}^-\right\rangle \to \Xi_c \bar D\left|{1}/{2}^- \right\rangle \gamma$ and $\Xi_c  \bar D^{*}\left|{1}/{2}^-\right\rangle \to \Xi_c \bar D\left|{1}/{2}^-\right\rangle \gamma$ processes \cite{Wang:2022tib}. Furthermore, the magnetic moments of the isoscalar $\Xi_c^{(\prime,*)} D^{(*)}$ molecular candidates can provide valuable information to reflect their inner structures and properties.

With more experimental data accumulation in Run II and Run III of LHCb and after the High-Luminosity-LHC upgrade \cite{Bediaga:2018lhg}, there is a good chance that LHCb will be able to detect the double-charm molecular pentaquarks in the near future. We strongly recommend that our experimental colleagues focus on the isoscalar $\Xi_c^{(\prime,*)} D^{(*)}$ molecular pentaquarks in the future, which not only enrich the family of the double-charm molecular states, but also test the interpretation of the observed $P_{\psi s}^{\Lambda}(4459)$, $P_{\psi s}^{\Lambda}(4338)$, and $T_{cc}(3875)^+$ states as the $\Xi_c \bar D^{*}$, $\Xi_c \bar D$, and $DD^{*}$ molecular states.

\section*{ACKNOWLEDGMENTS}

{To my mum (Mrs. Jun Shi) with this article.} This work is supported by the China National Funds for Distinguished Young Scientists under Grant No. 11825503, National Key Research and Development Program of China under Contract No. 2020YFA0406400, the 111 Project under Grant No. B20063, the fundamental Research Funds for the Central Universities, the project for top-notch innovative talents of Gansu province, and the National Natural Science Foundation of China under Grant Nos. 12335001, 12247155 and 12247101. F.L.W. is also supported by the China Postdoctoral Science Foundation under Grant No. 2022M721440.

\end{document}